\documentclass[smallcondensed]{svjour3}       
\journalname{Journal of Scientific Computing}
\usepackage{amsfonts}
\usepackage{latexsym}
\usepackage{yfonts}
\usepackage{amssymb}
\usepackage{mathrsfs}
\usepackage{upgreek}
\usepackage{bm}
\usepackage{dcolumn}
\usepackage{epsfig}
\usepackage{graphicx}
\usepackage{latexsym}
\usepackage{rotating}
\usepackage{soul}
\usepackage[english]{babel}

\usepackage{lmodern}
\usepackage[T1]{fontenc}
\usepackage[utf8]{inputenc}

\usepackage{times}
\usepackage{setspace}
\usepackage[mathcal]{euscript}

\usepackage{enumitem}

\usepackage{float}

\usepackage{setspace}
\usepackage[hang]{footmisc}
\usepackage{lipsum}

\usepackage{color}
\usepackage[colorlinks=true]{hyperref}

\AtBeginDocument{%
  \hypersetup{%
    linkcolor=[rgb]{0.1,0.4,0.1},%
    citecolor=[rgb]{0.6,0,0.1},%
    urlcolor=[rgb]{0.1,0,0.6}
  }%
}%

\DeclareTextFontCommand{\emph}{\sl}

\newcommand{\icol}[1]{
  \left[\begin{smallmatrix}#1\end{smallmatrix}\right]%
}

\usepackage{cite}
\soulregister\cite7
\soulregister\ref7
\soulregister\pageref7

\usepackage{amsmath}

\definecolor{mogreen}{rgb}{0.2,0.5,0.2}

\def\makeheadbox{\relax}

\begin{document}

\title{Particle-without-Particle: a practical pseudospectral collocation method for linear partial differential equations with distributional sources}

\titlerunning{Particle-without-Particle}

\author{Marius Oltean$\bm{^{1,2,3,4,5}}$          \and
        Carlos F. Sopuerta$\bm{^{1,2}}$
    \and
        Alessandro D.A.M. Spallicci$\bm{^{4,5,6,7}}$
}

\authorrunning{M. Oltean, C.F. Sopuerta and A.D.A.M. Spallicci}

\institute{Marius Oltean \at
        \email{oltean@ice.cat}  
    \and
        Carlos F. Sopuerta \at
        \email{sopuerta@ice.cat}
    \and
        Alessandro D.A.M. Spallicci \at
        \email{spallicci@cnrs-orleans.fr}
    \and
        $^{1}$ Institute of Space Sciences (ICE, CSIC), Campus Universitat Aut\`{o}noma de Barcelona,\\
        $^{\enskip}$ Carrer de Can Magrans s/n, 08193 Cerdanyola del Vall\`{e}s (Barcelona), Spain \vspace{0.15cm}\\
        $^{2}$ Institut d'Estudis Espacials de Catalunya (IEEC),\\
        $^{\enskip}$ Carrer del Gran Capità, 2-4, Edifici Nexus, despatx 201, 08034 Barcelona, Spain \vspace{0.15cm}\\
        $^{3}$ Departament de F\'{i}sica, Facultat de Ci\`{e}ncies, Universitat Aut\`{o}noma de Barcelona,\\
        $^{\enskip}$ Edifici C, 08193 Cerdanyola del Vall\`{e}s (Barcelona), Spain \vspace{0.15cm}\\
        $^{4}$ Observatoire des Sciences de l'Univers en r\'{e}gion Centre (OSUC), Universit\'{e} d'Orl\'{e}ans,\\
        $^{\enskip}$ 1A rue de la F\'{e}rollerie, 45071 Orl\'{e}ans, France \vspace{0.15cm}\\
        $^{5}$ Laboratoire de Physique et Chimie de l'Environnement et de l'Espace (LPC2E),\\
        $^{\enskip}$ Centre National de la Recherche Scientifique (CNRS),\\
        $^{\enskip}$ 3A Avenue de la Recherche Scientifique, 45071 Orl\'{e}ans, France \vspace{0.15cm}\\
        $^{6}$ P\^{o}le de Physique, Collegium Sciences et Techniques (CoST), Universit\'{e} d'Orl\'{e}ans,\\
        $^{\enskip}$ Rue de Chartres, 45100  Orl\'{e}ans, France \vspace{0.15cm}\\
        $^{7}$ Departamento de Astrof\'{\i}sica, Cosmologia e Intera\c{c}\~{o}es Fundamentais (COSMO),\\
        $^{\enskip}$ Centro Brasileiro de Pesquisas F\'{\i}sicas (CBPF),\\
        $^{\enskip}$ Rua Xavier Sigaud 150, 22290-180 Urca, Rio de Janeiro, Brazil
}

\date{}

\maketitle

\begin{abstract}
Partial differential equations with distributional sources---in particular, involving (derivatives of) delta distributions---have become increasingly ubiquitous in numerous areas of physics and applied mathematics. It is often of considerable interest to obtain numerical solutions for such equations, but any singular (``particle''-like) source modeling invariably introduces nontrivial computational obstacles. A common method to circumvent these is through some form of delta function approximation procedure on the computational grid; however, this often carries significant limitations on the efficiency of the numerical convergence rates, or sometimes even the resolvability of the problem at all.\\
In this paper, we present an alternative technique for tackling such equations which avoids the singular behavior entirely: the ``Particle-without-Particle'' method. Previously introduced in the context of the self-force problem in gravitational physics, the idea is to discretize the computational domain into two (or more) disjoint pseudospectral (Chebyshev-Lobatto) grids such that the ``particle'' is always at the interface between them; thus, one only needs to solve homogeneous equations in each domain, with the source effectively replaced by jump (boundary) conditions thereon. We prove here that this method yields solutions to any linear PDE the source of which is any linear combination of delta distributions and derivatives thereof supported on a one-dimensional subspace of the problem domain. We then implement it to numerically solve a variety of relevant PDEs: hyperbolic (with applications to neuroscience and acoustics), parabolic (with applications to finance), and elliptic. We generically obtain improved convergence rates relative to typical past implementations relying on delta function approximations.

\keywords{Pseudospectral methods \and Distributionally-sourced PDEs \and
Gravitational self-force \and Neural populations \and Price formation}
\end{abstract}

\section{Introduction}
\label{intro}

Mathematical models often have to resort---be it out of expediency
or mere ignorance---to deliberately idealized descriptions of their
contents. A common idealization across different fields of applied
mathematics is the use of the Dirac delta distribution, often simply
referred to as the \textsl{delta ``function''}, for the purpose of describing
highly localized phenomena: that is to say, phenomena the length
scale of which is significantly smaller, in some suitable sense, than that
of the problem into which they figure, and the (possibly complicated)
internal structure of which can thus be safely (or safely enough) ignored in
favour of a simple ``point-like'' cartoon. Canonical examples of this
from physics are notions such as ``point masses'' in gravitation or
``point charges'' in electromagnetism. 

Yet, despite their potentially powerful conceptual simplifications, introducing distributions into any mathematical
model is something that must be handled with
great technical care. In particular, let us suppose that our problem of interest
has the very general form
\begin{equation}
\mathcal{L}u=S\quad\mbox{in}\enskip\mathscr{U}\subseteq\mathbb{R}^{n}\,,\label{eq:intro_general_pde}
\end{equation}
where $\mathcal{L}$ is an $n$-dimensional (partial, if $n>1$) differential
operator (of arbitrary order $m$), $u$ is a quantity to be solved
for (a function, a tensor etc.) and we assume that $S$---the ``source''---is
distributional in nature, i.e. we have $S:\mathcal{D}(\mathscr{U})\rightarrow\mathbb{R}$, where we use the common notation $\mathcal{D}(\mathscr{U})$ to
refer to the set of smooth compactly-supported functions, i.e. ``test
functions'', on $\mathscr{U}$. It follows, therefore, that $u$---if
it exists---must also be distributional in nature. So strictly speaking,
from the point of view of the classic theory of distributions \cite{schwartz_theorie_1957}, the problem (\ref{eq:intro_general_pde})
is only well-defined---and hence may admit distributional solutions
$u$---provided that $\mathcal{L}$ is linear\footnote{Here the terms ``linear''/``nonlinear'' have their standard meaning from the theory of partial differential equations.}. 

The problem with a nonlinear
$\mathcal{L}$ is essentially that, classically, products of distributions
do not make sense \cite{schwartz_sur_1954}. While there has certainly been work by mathematicians
aiming to generalize the theory of distributions so as to accommodate
this possibility \cite{li_review_2007,colombeau_nonlinear_2013,bottazzi_grid_2017}, in the standard setting we are only really allowed
to talk of \textsl{linear} problems of the form (\ref{eq:intro_general_pde}).
Opportunely, very many of the typical problems in physics and applied
mathematics involving distributions take precisely this form. 

The inspiration for considering (\ref{eq:intro_general_pde}) in general
in this paper actually comes from a setting where one does, in fact,
encounter non-linearities a priori: namely, gravitational physics. (For a general discussion regarding the treatment of distributions therein, see Ref. \cite{geroch_strings_1987}.)
In particular, equations such as (\ref{eq:intro_general_pde}) arise
when attempting to describe the backreaction of a body with a ``small''
mass upon the spacetime through which it moves---known as its \textsl{self-force} \cite{mino_gravitational_1997,quinn_axiomatic_1997,detweiler_self-force_2003,gralla_rigorous_2008,gralla_note_2011,poisson_motion_2011,blanchet_mass_2011,spallicci_self-force_2014,pound_motion_2015,wardell_self-force:_2015}.
(A similar version of this problem exists in electromagnetism, where
a ``small'' charge backreacts upon the electromagnetic field that
determines its motion \cite{dirac_classical_1938,dewitt_radiation_1960,barut_electrodynamics_1980,poisson_motion_2011}.) In the full Einstein equations of general
relativity, which can be regarded as having the schematic form (\ref{eq:intro_general_pde})
with $u$ describing the gravitational field (that is, the spacetime
geometry, in the form of the metric) and $S$ denoting the matter
source (the stress-energy-momentum tensor), $\mathcal{L}$ is a nonlinear
operator. Nevertheless, for a distributional $S$ (representing the
``small'' mass as a ``point particle'' source) one \textsl{can} legitimately
seek solutions to a \textsl{linearized} version of (\ref{eq:intro_general_pde})
in the context of perturbation theory, i.e. at first order in an expansion
of $\mathcal{L}$ in the mass. The detailed problem, in this case,
turns out to be highly complex, and in practice, $u$ must be computed
numerically. The motivation for this, we may add, is not just out of purely theoretical or foundational concern---the calculation of the self-force is also of significant applicational value for gravitational wave astronomy. To wit, it will in fact be indispensable for generating accurate enough waveform templates for future space-based gravitational wave detectors such as LISA \cite{amaro-seoane_et_al._gravitational_2013,amaro-seoane_et_al._laser_2017} vis-à-vis extreme-mass-ratio binary systems, which are expected to be among the most fruitful sources thereof. For these reasons, having at our disposal a practical and
efficient numerical method for handling equations of the form (\ref{eq:intro_general_pde})
is of consequential interest.

What is more, these sorts of partial differential equations (PDEs)
arise frequently in other fields as well; indeed, (\ref{eq:intro_general_pde})
can adequately characterize quite a wide variety of (linear) mathematical
phenomena assumed to be driven by ``localized sources''. A few examples,
which we will consider one by one in different sections of this paper,
are the following:

\begin{enumerate}[label=(\roman*)]

\item \textsl{First-order hyperbolic PDEs}: in neuroscience, advection-type
PDEs with a delta function source can be used in the modeling of neural
populations \cite{haskell_population_2001,casti_population_2002,caceres_analysis_2011,caceres_blow-up_2016};

\item \textsl{Parabolic PDEs}: in finance, heat-type PDEs with delta
function sources are sometimes used to model price formation \cite{lasry_mean_2007,markowich_parabolic_2009,caffarelli_price_2011,burger_boltzmann-type_2013,achdou_partial_2014,pietschmann_partial_2012};

\item \textsl{Second-order hyperbolic PDEs}: in acoustics, wave-type
PDEs with delta function (or delta derivative) sources are used to
model monopoles (or, respectively, multipoles) \cite{petersson_stable_2010,kaltenbacher_computational_2017}; more complicated equations of this form also appear, for example, in seismology models \cite{romanowicz_seismology_2007,aki_quantitative_2009,shearer_introduction_2009,madariaga_seismic_2007,petersson_stable_2010}, which we will briefly comment upon.

\item \textsl{Elliptic PDEs}: Finally, we will look at a simple Poisson equation with a singular source \cite{tornberg_numerical_2004}; such equations can describe, for example, the potential produced by a very localized charge in electrostatics.

\end{enumerate}

\subsection{Scope of this paper}

The purpose of this paper is to explicate and generalize a practical method for numerically solving equations
like (\ref{eq:intro_general_pde}), as well as to illustrate its broad
applicability to the various problems listed in (i)-(iv) above. Previously
implemented with success only in the specific context of the self-force
problem \cite{canizares_extreme-mass-ratio_2011,canizares_efficient_2009,canizares_pseudospectral_2010,canizares_time-domain_2011,canizares_tuning_2011,jaramillo_are_2011,canizares_overcoming_2014,oltean_frequency-domain_2017}, we dub it the ``Particle-without-Particle'' (PwP) method.
(Other methods for the computation of the self-force have also been developed based on matching the properties of the solutions on the sides of the delta distributions---see, e.g., the indirect (source-free) integration method of Refs. \cite{aoudia_source-free_2011,ritter_fourth-order_2011,spallicci_towards_2012,spallicci_fully_2014,ritter_indirect_2015,ritter_indirect_2015-1}.)
The basic idea of the PwP approach is the following: One begins by writing $u$ as a sum of distributions
each of which has support outside (plus, if necessary, at the location
of) the points where $S$ is supported; one then solves the equations
for each of these pieces of $u$ and finally matches them in such
a way that their sum satisfies the original problem (\ref{eq:intro_general_pde}).
In fact, as we shall soon elaborate upon, this approach will not work
in general for all possible problems of the form (\ref{eq:intro_general_pde}).
However, we will prove that it will \textsl{always} work if, rather
than the source being a distribution defined on all of $\mathscr{U}$,
we have instead $S:\mathcal{D}(\mathscr{I})\rightarrow\mathbb{R}$
with $\mathscr{I}\subseteq\mathbb{R}$ representing a \textsl{one-dimensional subspace}
of $\mathscr{U}$.

To make things more concrete, let us briefly describe this procedure
using the simplest possible example: let $f:\mathscr{U}\rightarrow\mathbb{R}$
be an arbitrary given function and suppose $S=f\delta$ where $\delta:\mathcal{D}(\mathscr{I})\rightarrow\mathbb{R}$
is the delta function supported at some point $x_{p}\in\mathscr{I}$.
Then, to solve (\ref{eq:intro_general_pde}), one would assume the
decomposition (or ``ansatz'') $u=u^{-}\Theta^{-}+u^{+}\Theta^{+}$ with
$\Theta^{\pm}:\mathcal{D}(\mathscr{I})\rightarrow\mathbb{R}$ denoting
appropriately defined Heaviside distributions (supported to the right/left
of $x_{p}$, respectively), and $u^{\pm}:\mathscr{U}\rightarrow\mathbb{R}$ being simple functions
(not distributions) to be solved for. Inserting such a decomposition for $u$ into
(\ref{eq:intro_general_pde}), one obtains \textsl{homogeneous} equations
$\mathcal{L}u^{\pm}=0$ on the appropriate domains, supplemented by
the necessary boundary conditions (BCs) for these equations at $x_{p}\in\mathscr{I}$,
explicitly determined by $f$. Generically, the latter arise in the
form of relations between the limits of $u^{-}$ and $u^{+}$ (and/or
the derivatives thereof) at $x_{p}$, and for this reason are called
``jump conditions'' (JCs). Effectively, the latter completely replace
the ``point'' source $S$ in the original problem, now simply reduced
to solving sourceless equations---hence the nomenclature of the method.

While in principle one can certainly contemplate the adaptation of these ideas into a variety of established approaches for the numerical solution of PDEs, we will focus specifically on their implementation through pseudospectral collocation (PSC) methods on Chebyshev-Lobatto (CL) grids. The principal advantages thereof lie in their typically very efficient (exponential) rates of numerical convergence as well as the ease of incorporating and modifying BCs (JCs) throughout the evolution. Indeed, PSC methods have enjoyed very good success in past work \cite{canizares_extreme-mass-ratio_2011,canizares_efficient_2009,canizares_pseudospectral_2010,canizares_time-domain_2011,canizares_tuning_2011,jaramillo_are_2011,canizares_overcoming_2014,oltean_frequency-domain_2017} on the PwP approach for self-force calculations (and in gravitational physics more generally \cite{grandclement_spectral_2009}, including arbitrary precision implementations \cite{santos-olivan_pseudo-spectral_2018}), and so we shall not deviate very much from this recipe in the models considered in this paper. Essentially the main difference will be that here, instead of the method of lines which featured in most of the past PwP self-force work, we will for the most part carry out the time evolution using the simplest first-order forward finite difference scheme; we do this, on the one hand, so that we may illustrate the principle of the method explicitly in a very elementary way without too many technical complications, and on the other, to show how well it can work even with such basic tactics. Depending on the level of accuracy and computational efficiency required for any realistic application, these procedures can naturally be complexified (to higher order, more domains, more complicated domain compactifications etc.) for properly dealing with the sophistication of the problem at hand.

To summarize, past work using the PwP method only solved a specific
form of Eq. (\ref{eq:intro_general_pde}) pertinent to the self-force
problem: that is, with a particular choice of $\mathcal{L}$ and $S$
(upon which we will comment more later). It did \textit{not} consider
the question of the extent to which the idea of the method could be
useful in general for solving distributionally-sourced PDEs. These
appear, as enumerated above, in many other fields of study---and
we submit that a method such as this would be of valuable benefit
to researchers working therein. The novelty of the present paper will
thus be to formulate a \textit{completely general} PwP method for
\textit{any} distributionally-sourced (linear) problem of the form
(\ref{eq:intro_general_pde}) with the single limiting condition that
${\rm supp}(S)\subset\mathscr{I}\subseteq\mathscr{U}$ where ${\rm dim}(\mathscr{I})=1$.
We will prove rigorously why and how the method works for such problems,
and then we will implement it to obtain numerical solutions to the
variety of different applications mentioned earlier in order to illustrate
its broad practicability. We will see that, in general, this method
either matches or improves upon the results of other methods existent in the literature
for tackling distributionally-sourced PDEs---and we turn to a more
detailed discussion of this topic in the next subsection.

\subsection{Comparison with other methods in the literature}

Across all areas of application, the most commonly encountered---and,
perhaps, most naively suggestible---strategy for numerically solving
equations of the form (\ref{eq:intro_general_pde}) is to rely upon
some sort of delta function approximation procedure on the computational
grid \cite{tornberg_numerical_2004,jung_collocation_2009,jung_note_2009,petersson_discretizing_2016}.
For instance, the simplest imaginable choice in this vein is just
a narrow hat function (centered at the point where the delta function
is supported, and having total measure $1$) which, for better accuracy,
one can upgrade to higher-order polynomials, or even trigonometric
functions. Another readily evocable possibility is to use a narrow
Gaussian---and indeed, this is one option that has in fact been tried
in self-force computations as well (see Ref. \cite{lopez-aleman_perturbative_2003},
for example). However, this unavoidably introduces
into the problem an additional, artificial length scale: that is,
the width of the Gaussian, which a priori need not have anything to
do with the actual (\textquotedblleft physical\textquotedblright )
length scale of the source. Moreover, there is the evident drawback
that no matter how small this artificial length scale is chosen, the
solutions will never be well-resolved close to the distributional
source location: there will always be some sort of Gibbs-type phenomenon\footnote{The Gibbs phenomenon, originally discovered by Henry Wilbraham \cite{wilbraham_certain_1848} and rediscovered by J. Willard Gibbs \cite{gibbs_letter_1899}, refers generally to an overshoot in the approximation of a piecewise continuously differentiable function near a jump discontinuity.}
there.

Methods for solving (\ref{eq:intro_general_pde}) which are closer
in spirit to our PwP method have been explored in Refs. \cite{field_discontinuous_2009}
and \cite{shin_spectral_2011}. In particular, both of these works
have used the idea of placing the distributional source at the interface
of computational grids---however, they tackle the numerical implementation
differently than we do.

In the case of Ref. \cite{field_discontinuous_2009}---which, incidentally,
is also concerned with the self-force problem---the difference is
that the authors use a discontinuous Galerkin method (rather than
spectral methods, as in our PwP approach), and the effect of the distributional
source is accounted for via a modification of the numerical flux at
the ``particle'' location. This relies essentially upon a weak formulation
of the problem, wherein a choice has to be made about how to assign
measures to the distributional terms over the relevant computational
domains. In contrast, we directly solve only for smooth solutions
supported away from the ``particle'' location, and account for the
distributional source simply by imposing adequate boundary---i.e. jump---
conditions there.

Ref. \cite{shin_spectral_2011} is closer to our approach in this
sense, as the authors there also use spectral methods and also account
for the distributional source via jump conditions. However, the difference
with our method is that Ref. \cite{shin_spectral_2011} treats these
jump conditions as additional constraints (rather than built-in boundary
conditions) for the smooth solutions away from the distributional
source, thus over-determining the problem. That being the case, the
authors are led to the need to define a functional (expressing how
well the differential equations plus the jump conditions are satisfied)
to be minimized---constituting what they refer to as a ``least squares
spectral collocation method''. There is however no unique way to
choose this functional. Moreover, the complication of introducing
it is not at all necessary: our approach, in contrast, simply replaces
the discretization of (the homogeneous version of) the differential
equations at the ``particle'' location with the corresponding jump
conditions (i.e. it imposes the jump conditions \textit{as} boundary
conditions, \textit{by construction}---something which PSC methods are precisely designed to be able to handle), leading to completely determined
systems in all cases which are solved directly, without further complications.

Finally, neither Ref. \cite{field_discontinuous_2009} nor \cite{shin_spectral_2011}
analyzed to any significant extent the conditions under which their
methods might be applicable to more general distributionally-sourced
PDEs. As mentioned, in the present paper we will devote a careful
proof entirely to this issue.

This paper is structured as follows. Following some mathematical preliminaries
in Section \ref{setup}, we prove in Section \ref{pwp} how the PwP method can be formulated and applied to problems with the most general possible ``point'' source $S:\mathcal{D}(\mathscr{I})\rightarrow\mathbb{R}$,
that is, one containing an arbitrary number of (linearly combined) delta
derivatives and supported at an arbitrary number of points in $\mathscr{I}$.
Thus, one can use it on any type of (linear) PDE involving such sources,
which we illustrate with the applications listed in (i)-(iv) above
in Sections \ref{first-order-hyperbolic-pdes}-\ref{elliptic-pdes}
respectively. Finally, we give concluding remarks in Section \ref{conclusions}.

\section{Setup}
\label{setup}

We wish to begin by establishing some basic notation and then reviewing
some pertinent properties of distributions that we will need to make
use of later on. While we will certainly strive to maintain a fair level of mathematical
rigour here and throughout this paper (at least, insofar as a certain
amount of formal precaution is inevitably necessary when dealing
with distributions), our principal aim remains that of presenting
practical methodologies; hence the word ``distribution'' may at times be liberally interchanged for
``function'' (e.g. we may say ``delta function'' instead of ``delta distribution'') and some notation possibly
slightly abused, when the context is clear enough to not pose dangers
for confusion.

\subsection{Distributionally-sourced linear PDEs}

Consider the problem (\ref{eq:intro_general_pde}) with $S:\mathcal{D}(\mathscr{I})\rightarrow\mathbb{R}$,
where $\mathscr{I}\subseteq\mathbb{R}$ is a one-dimensional subspace
of $\mathscr{U}\subseteq\mathbb{R}^{n}$, as discussed in the introduction.
Then we can view $\mathscr{U}$ as a product space, $\mathscr{U}=\mathscr{I}\times\mathscr{V}$
with $\mathscr{V}=\mathscr{U}/\mathscr{I}\subseteq\mathbb{R}^{n-1}$,
and write coordinates on $\mathscr{U}$ as $\mathbf{x}=(x,\mathbf{y})$
with $x\in\mathscr{I}$ and $\mathbf{y}=(y_{1},\ldots,y_{n-1})\in\mathscr{V}$,
such that
\begin{align}
f:\mathscr{U}=\mathscr{I}\times\mathscr{V}\subseteq\mathbb{R}\times\mathbb{R}^{n-1}=\mathbb{R}^{n} & \rightarrow\mathbb{R}\nonumber\\
\mathbf{x}\,\,=\,\,\left(x,\mathbf{y}\right)\,=\,\left(x,y_{1},\ldots,y_{n-1}\right) & \mapsto f\left(\mathbf{x}\right)\label{eq:function_general} 
\end{align}
denotes any arbitrary function on $\mathscr{U}$.

It is certainly possible, in the setup we are about to describe, to
have $\mathscr{V}=\emptyset$, i.e. problems involving just \emph{ODEs}
(on $\mathscr{U}=\mathscr{I}$) of the form (\ref{eq:intro_general_pde})---and,
in fact, our first elementary example illustrating the PwP method
in the following section will be of such a kind. For the more involved
numerical examples we will study in later sections, we will most often
be dealing with functions of two variables, $x\in\mathscr{I}$ for
``space'' (or some other pertinent parameter) and $t\in\mathscr{V}\subseteq\mathbb{R}$
for time.

For any function (\ref{eq:function_general}) involved in these problems,
we will sometimes use the notation $f'=\partial_{x}f$ for the ``spatial''
derivative; also, we may employ $\dot{f}=\partial_{t}f$ for the partial
derivative with respect to time $t$ when $\{t\}$ is (a subspace
of) $\mathscr{V}$.

Now, as in the introduction, let $\mathcal{L}$ be any general $m$-th
order \emph{linear} differential operator. The sorts of PDEs (\ref{eq:intro_general_pde})
that we will be concerned with have the basic form 
\begin{equation}
\mathcal{L}u=S=f\delta_{\left(p\right)}+g\delta_{\left(p\right)}^{'}+\cdots\,,\label{eq:setup-pde_general_1}
\end{equation}
where $f(\mathbf{x})$, $g(\mathbf{x})$ etc. are ``source'' functions
prescribed by the problem at hand, and we employ the convenient notation
\begin{equation}
\delta_{\left(p\right)}\left(x\right)=\delta\left(x-x_{p}\left(\mathbf{y}\right)\right)
\end{equation}
to indicate the Dirac delta distribution on $\mathscr{I}$ centered
at the ``particle location'' $x_{p}(\mathbf{y})$—the functional
form of which can be either specified a priori, or determined via
some given prescription as the solution $u$ itself is evolved. When
there is no risk of confusion, we may sometimes omit the $\mathbf{y}$
dependence in our notation and simply write $x_{p}$.

In fact, our PwP method can even deal with multiple, say $M$, ``particles''. PwP computations of the self-force have actually only required $M=1$ (there being only one ``particle'' involved in the problem), so the general $M\geq1$ case has not been considered up to now.
Concordantly, to express our problem of interest (\ref{eq:setup-pde_general_1})
in the most general possible form, let us employ the typical PDE notation
for ``multi-indices'' \cite{evans_partial_1998}, $\alpha=(\alpha_{0},\alpha_{1},\ldots,\alpha_{n-1})$
with each $\alpha_{I}\in\mathbb{Z}^{\geq}$ being a non-negative integer (indexed from $I=0$ to $I=n-1$ so as to make sense vis-à-vis our coordinate notation on $\mathscr{U}$, instead of the more usual practice to label them from $1$ to $n$),
and $|\alpha|=\sum_{I=0}^{n-1}\alpha_{I}$. Furthermore, we define $\alpha!=\prod_{I=0}^{n-1}\alpha_{I}!$. Thus, the most general
$m$-th order linear partial differential operator can be written
as $\mathcal{L}=\sum_{|\alpha|\leq m}\xi^{\alpha}(\mathbf{x})D^{\alpha}$
where $\xi^{\alpha}:\mathscr{U}\rightarrow\mathbb{R}$ are arbitrary functions and $D^{\alpha}=\partial^{|\alpha|}/\partial x^{\alpha_{0}}\partial y_{1}^{\alpha_{1}}\cdots\partial y_{n-1}^{\alpha_{n-1}}$.
Hence, we are dealing with any problem which can be placed into the
form 
\begin{equation}
\sum_{|\alpha|\leq m}\xi^{\alpha}\left(\mathbf{x}\right)D^{\alpha}u\left(\mathbf{x}\right)=\sum_{i=1}^{M}\sum_{j=0}^{K}f^{ij}\left(\mathbf{x}\right)\delta^{(j)}\left(x-x_{p_{i}}\left(\mathbf{y}\right)\right)\,,\label{eq:setup-pde_general_2}
\end{equation}
with $f^{ij}:\mathscr{U}\rightarrow\mathbb{R}$ denoting the ``source''
functions (for the $j$-th delta derivative of the $i$-th particle)
and $K\in\mathbb{Z}^{\geq}$ the highest order of the delta function
derivatives in $S$, appropriately supplemented by initial/boundary
conditions (ICs/BCs).

Let us give a few basic examples to render this setup more palpable.
One very simple example---that which will serve as our first illustration
of the PwP method in the next section---is the simple harmonic oscillator
with a constant delta function forcing (source) term---that is, the
ODE (with $\mathscr{V}=\emptyset$): 
\begin{equation}
u''+u=a\delta_{\left(p\right)}\,,
\end{equation}
where $\delta_{\left(p\right)}(x)=\delta(x-x_{p})$ for some fixed
$x_{p}\in\mathscr{I}$, and $a\in\mathbb{R}$. Another example is
the wave equation with a moving singular source, 
\begin{equation}
\left(\partial_{t}^{2}-\partial_{x}^{2}\right)u\left(x,t\right)=f\left(x,t\right)\delta\left(x-x_{p}\left(t\right)\right)\,,
\end{equation}
with $x_{p}(t)$ specified as a function of time.

\subsection{Properties of distributions}

We now wish to remind the reader of a few basic properties of distributions
before proceeding to describe the PwP procedure; for a good detailed exposition, see e.g. Ref. \cite{stakgold_greens_2011}.

Let $f:\mathscr{U}\rightarrow\mathbb{R}$ be, as before, any function
involved in the problem (\ref{eq:setup-pde_general_2}). We denote
by
\begin{align}
f_{p}:\mathscr{V}\rightarrow\, & \mathbb{R}\nonumber \\
\mathbf{y}\mapsto\, & f_{p}\left({\bf y}\right)=f\left(x_{p}\left({\bf y}\right),{\bf y}\right)
\end{align}
the function evaluated at the ``particle'' position.

Furthermore, let $\phi\in\mathcal{D}(\mathscr{I})$ be any test function
on $\mathscr{I}$. Then we define the action of the distribution associated
with $f$ as: 
\begin{equation}
\left\langle f,\phi\right\rangle =\int_{\mathscr{I}}{\rm d}x\,f\left(x,\mathbf{y}\right)\phi\left(x\right)\,.
\end{equation}
We say that two functions $f$ and $g$ are \emph{equivalent} in the
sense of distributions if 
\begin{equation}
\langle f,\phi\rangle=\langle g,\phi\rangle\Leftrightarrow f\equiv g\,.
\end{equation}

An identity which will be important for us in discussing the PwP method
is the following \cite{cortizo_diracs_1995,li_review_2007}: 
\begin{equation}
f\left(x,\mathbf{y}\right)\delta_{\left(p\right)}^{\left(n\right)}\left(x\right)\equiv\left(-1\right)^{n}\sum_{j=0}^{n}\left(-1\right)^{j}\binom{n}{j}f_{p}^{\left(n-j\right)}\left(\mathbf{y}\right)\delta_{\left(p\right)}^{\left(j\right)}\left(x\right)\,,\label{eq:setup-distributions_identity}
\end{equation}
where $\delta_{\left(p\right)}^{\left(n\right)}=\partial_{x}^{n}\delta_{\left(p\right)}$.
For concreteness, let us write down the first three cases explicitly
here: 
\begin{align}
f\left(x,\mathbf{y}\right)\delta_{\left(p\right)}\left(x\right)\equiv\, & f_{p}\left(\mathbf{y}\right)\delta_{\left(p\right)}\left(x\right)\,,\\
f\left(x,\mathbf{y}\right)\delta_{\left(p\right)}'\left(x\right)\equiv\, & -f_{p}'\left(\mathbf{y}\right)\delta_{\left(p\right)}\left(x\right)+f_{p}\left(\mathbf{y}\right)\delta_{\left(p\right)}'\left(x\right)\,,\\
f\left(x,\mathbf{y}\right)\delta_{\left(p\right)}''\left(x\right)\equiv\, & f_{p}''\left(\mathbf{y}\right)\delta_{\left(p\right)}\left(x\right)-2f_{p}'\left(\mathbf{y}\right)\delta_{\left(p\right)}'\left(x\right)+f_{p}\left(\mathbf{y}\right)\delta_{\left(p\right)}''\left(x\right)\,.
\end{align}
For the interested reader, we offer in Appendix \ref{a-proof-identity}
a proof by induction of the formula (\ref{eq:setup-distributions_identity}),
which is instructive for appreciating the subtleties generally involved
in manipulating distributions.

Let 
\begin{equation}
\Theta_{\left(p\right)}^{\pm}\left(x\right)=\Theta\left(\pm\left(x-x_{p}\left(\mathbf{y}\right)\right)\right)
\end{equation}
be the Heaviside function which is supported to the right/left (respectively)
of $x_{p}$. Then, we have: 
\begin{align}
\partial_{x}\Theta_{\left(p\right)}^{\pm}=\, & \pm\delta_{\left(p\right)}\,,\label{eq:setip-partial_x_Theta}\\
\partial_{y_{j}}\Theta_{\left(p\right)}^{\pm}=\, & \mp\left(\partial_{y_{j}}x_{p}\right)\delta_{\left(p\right)}\,,\label{eq:setup-partial_y_Theta}
\end{align}
and so on for higher order partials.

For notational expediency, we may sometimes omit the $(p)$ subscript
on the Heaviside functions (and derivatives thereof) when the context
is sufficiently clear.

\section{The ``Particle-without-Particle'' method}
\label{pwp}

As discussed heuristically in the introduction, the basic idea of
our method for solving (\ref{eq:setup-pde_general_2}) is to effectively
eliminate the ``point''-like source or ``particle'' from the problem
by decomposing the solution $u$ into a series of distributions: specifically,
Heaviside functions $\Theta^{i}:\mathcal{D}(\mathscr{I})\rightarrow\mathbb{R}$
supported in each of the $M+1$ disjoint regions of $\mathscr{I}\backslash{\rm supp}(S)$
(i.e. ${\rm supp}(\Theta^{i})\cap{\rm supp}(S)=\emptyset,\forall i$
and ${\rm supp}(\Theta^{i})\cap{\rm supp}(\Theta^{j})=\emptyset,\forall i\neq j$)
and, if necessary, delta functions (plus delta derivatives) at ${\rm supp}(S)$:
\begin{equation}
u=\sum_{i=0}^{M}u^{i}\Theta^{i}+\sum_{i=1}^{M}\sum_{j=0}^{K-m}h^{ij}\delta_{\left(p_{i}\right)}^{\left(j\right)}\,,\label{eq:pwp-decomposition}
\end{equation}
where $u^{i}:\mathscr{U}\rightarrow\mathbb{R}$ and we need to include
the second sum with $h^{ij}:\mathscr{V}\rightarrow\mathbb{R}$ only
if $K\geq m$.

We will prove in this section that one can always obtain solutions
of the form (\ref{eq:pwp-decomposition}) to the problem (\ref{eq:setup-pde_general_2}).
In particular, inserting (\ref{eq:pwp-decomposition}) into (\ref{eq:setup-pde_general_2})
will always yield homogeneous equations
\begin{equation}
\mathcal{L}u^{i}=0\quad\mbox{in}\,\,\left(\mathscr{I}\backslash{\rm supp}(S)\right)\times\mathscr{V}\,,
\end{equation}
along with JCs on (the derivatives of) $u$---and possibly (derivatives
of) $h^{ij}$ if applicable. In general, we define the ``jump'' $[\cdot]_{p}:\mathscr{V}\rightarrow\mathbb{R}$
in the value of any function $f:\mathscr{U}\rightarrow\mathbb{R}$
at $x_{p}({\bf y})$ as
\begin{equation}
\left[f\right]_{p}\left({\bf y}\right)=\lim_{x\rightarrow x_{p}({\bf y})^{+}}f\left(x,{\bf y}\right)-\lim_{x\rightarrow x_{p}({\bf y})^{-}}f\left(x,{\bf y}\right)\,.\label{eq:pwp-jump}
\end{equation}
Henceforth, for convenience, we will generally omit the ${\bf y}$-dependence
and simply write $[f]_{p}$.

First we will work through a simple example in order to offer a more
concrete sense of the method, and afterwards we will show in general
how (\ref{eq:pwp-decomposition}) solves (\ref{eq:setup-pde_general_2}).

\subsection{Simple example}

We illustrate here the application of our PwP method to a very simple
ODE (and single-particle) example. We will consider the problem 
\begin{equation}
\mathcal{L}u=u''+u=a\delta+b\delta^{'}\,,\enskip x\in\mathscr{I}=\left[-L,L\right]\,,\enskip u\left(\pm L\right)=0\,,\label{eq:pwp-simple_example}
\end{equation}
where $\delta$ is simply the delta function centered at $x_{p}=0$.

We begin by decomposing $u$ as 
\begin{equation}
u=u^{-}\Theta^{-}+u^{+}\Theta^{+}\,,
\end{equation}
where $\Theta^{\pm}(x)=\Theta(\pm x)$, and we insert this into (\ref{eq:pwp-simple_example}).
Using (\ref{eq:setip-partial_x_Theta}), the LHS becomes simply 
\begin{align}
\mathcal{L}u\left(x\right)=\, & u''\left(x\right)+u\left(x\right)\\
=\, & \left\{ \mathcal{L}u^{-}\left(x\right)\right\} \Theta^{-}\left(x\right)+\left\{ \mathcal{L}u^{+}\left(x\right)\right\} \Theta^{+}\left(x\right)\nonumber \\
 & +\left\{ -2\left(u^{-}\left(x\right)\right)'+2\left(u^{+}\left(x\right)\right)'\right\} \delta\left(x\right)\nonumber \\
 & +\left\{ -u^{-}\left(x\right)+u^{+}\left(x\right)\right\} \delta'\left(x\right)\,.
\end{align}
Now before we can equate this to the distributional terms in the source (RHS), we must apply the identity (\ref{eq:setup-distributions_identity}).
In particular, we use $f(x)\delta(x)\equiv f_{p}\delta(x)$ and
$f(x)\delta'(x)\equiv-f_{p}'\delta(x)+f_{p}\delta'(x)$. Thus, the
above becomes 
\begin{align}
\mathcal{L}u\left(x\right)\equiv\, & \left\{ \mathcal{L}u^{-}\left(x\right)\right\} \Theta^{-}\left(x\right)+\left\{ \mathcal{L}u^{+}\left(x\right)\right\} \Theta^{+}\left(x\right)\nonumber \\
 & +\left\{ -2\left(u^{-}\right)'_{p}+2\left(u^{+}\right)'_{p}\right\} \delta\left(x\right)\nonumber \\
 & +\left\{ \left(u^{-}\right)'_{p}-\left(u^{+}\right)'_{p}\right\} \delta\left(x\right)+\left\{ -u_{p}^{-}+u_{p}^{+}\right\} \delta'\left(x\right)\\
=\, & \left\{ \mathcal{L}u^{-}\left(x\right)\right\} \Theta^{-}\left(x\right)+\left\{ \mathcal{L}u^{+}\left(x\right)\right\} \Theta^{+}\left(x\right)+\left[u'\right]_{p}\delta\left(x\right)+\left[u\right]_{p}\delta'\left(x\right)\,.
\end{align}
Plugging this into the DE (\ref{eq:pwp-simple_example}), we have
\begin{equation}
\left\{ \mathcal{L}u^{-}\right\} \Theta^{-}+\left\{ \mathcal{L}u^{+}\right\} \Theta^{+}+\left[u'\right]_{p}\delta+\left[u\right]_{p}\delta'\equiv a\delta+b\delta^{'}\,.
\end{equation}
Therefore the original problem is equivalent to the system of equations:
\begin{equation}
\begin{cases}
\mathcal{L}u^{-}=0, & x\in\mathscr{D}^{-}=\left[-L,0\right]\,,\enskip u^{-}\left(-L\right)=0\,,\\
\mathcal{L}u^{+}=0, & x\in\mathscr{D}^{+}=\left[0,L\right]\,,\enskip u^{+}\left(L\right)=0\,,\\
\left[u\right]_{p}=b, & \left[u'\right]_{p}=a\,.
\end{cases}\label{eq:pwp-simple_example_pwp}
\end{equation}

Let us solve (\ref{eq:pwp-simple_example_pwp}), for simplicity, taking
$L=\pi/4$. The left homogeneous equation in (\ref{eq:pwp-simple_example_pwp})
has the general solution $u^{-}=A^{-}\cos(x)+B^{-}\sin(x)$, and the
BC tells us that $0=u^{-}(-\pi/4)=\frac{1}{\sqrt{2}}(A^{-}-B^{-})$,
i.e. 
\begin{equation}
A^{-}-B^{-}=0\,.\label{eq:pwp-simple_example_constants1}
\end{equation}
The right homogeneous equation in (\ref{eq:pwp-simple_example_pwp})
similarly has general solution $u^{+}=A^{+}\cos(x)+B^{+}\sin(x)$,
with the BC stating $0=u^{+}(\pi/4)=\frac{1}{\sqrt{2}}(A^{+}+B^{+})$,
i.e. 
\begin{equation}
A^{+}+B^{+}=0\,.\label{eq:pwp-simple_example_constants2}
\end{equation}
So far we have two equations (\ref{eq:pwp-simple_example_constants1})-(\ref{eq:pwp-simple_example_constants2})
for four unknowns (the integration constants in the general solutions).
It is the JCs in (\ref{eq:pwp-simple_example_pwp}) that provide us
with the remaining necessary equations to fix the solution. We have
$u^{-}(0)=A^{-}$, $(u^{-})'(0)=B^{-}$, $u^{+}(0)=A^{+}$ and $(u^{+})'(0)=B^{+}$
(understood in the appropriate limit approaching $x_{p}=0$). Hence
the JCs tell us: 
\begin{align}
b=\, & \left[u\right]_{p}=u^{+}(0)-u^{-}(0)=A^{+}-A^{-}\,,\label{eq:pwp-simple_example_constants3}\\
a=\, & \left[u'\right]_{p}=(u^{+})'(0)-(u^{-})'(0)=B^{+}-B^{-}\,.\label{eq:pwp-simple_example_constants4}
\end{align}
(We can think of the JCs as a mixing of the degrees of freedom in
the homogeneous solutions in such a way that they ``link together''
to produce the solution generated by the original distributional source.)
Solving (\ref{eq:pwp-simple_example_constants1})-(\ref{eq:pwp-simple_example_constants4}),
we get $A^{-}=-\frac{a+b}{2}=B^{-}$, $A^{+}=-\frac{a-b}{2}=-B^{+}$.
We now have the full solution to our original problem (\ref{eq:pwp-simple_example}):
\begin{equation}
u\left(x\right)=-\frac{a+b}{2}\left(\cos\left(x\right)+\sin\left(x\right)\right)\Theta\left(-x\right)-\frac{a-b}{2}\left(\cos\left(x\right)-\sin\left(x\right)\right)\Theta\left(x\right)\,.
\end{equation}

\subsection{General proof}

Suppose we have $M$ ``particles'' located at ${\rm supp}(S)=\{x_{p_{i}}\}_{i=1}^{M}\subset\mathscr{I}$,
as in the problem (\ref{eq:setup-pde_general_2}), with $x_{p_{1}}<x_{p_{2}}<\cdots<x_{p_{M}}$.
(NB: For $M\geq2$, if there exists any subset of $\mathscr{V}$ where
it should happen that $x_{p_{i}}({\bf y})>x_{p_{i+1}}({\bf y})$ as
a consequence of the ${\bf y}$-evolution, we can, without loss of
generality, simply swap indices within that subset so as to always
have $x_{p_{i}}<x_{p_{i+1}},\forall i$.) Furthermore let us assume
for the moment that the maximum order of delta function derivatives
in the source is one less than the order of the PDE (or smaller),
i.e. $K=m-1$. In this case, we do not need to consider the second
term on the RHS of (\ref{eq:pwp-decomposition}), i.e. $u$ is just
split up into pieces which are supported only in between all the particle
locations: $u^{0}(\mathbf{x})$ to the left of $x_{p_{1}}$, $u^{1}(\mathbf{x})$
between $x_{p_{1}}$ and $x_{p_{2}}$, ..., $u^{i}(\mathbf{x})$ between
$x_{p_{i}}$ and $x_{p_{i+1}}$, ..., and finally $u^{M}$ to the
right of $x_{p_{M}}$. Thus, we take 
\begin{equation}
u=\sum_{i=0}^{M}u^{i}\Theta^{i}\,,\label{eq:pwp-decomposition_general}
\end{equation}
where we define
\begin{equation}
\Theta^{i}=\begin{cases}
\Theta_{\left(p_{1}\right)}^{-}\,, & i=0\,,\\
\Theta_{\left(p_{i}\right)}^{+}-\Theta_{\left(p_{i+1}\right)}^{+}\,, & 1\leq i\leq M-1\,,\\
\Theta_{\left(p_{M}\right)}^{+}\,, & i=M\,,
\end{cases}
\end{equation}
denoting, as before, $\Theta_{(p_{i})}^{\pm}(x)=\Theta(\pm(x-x_{p_{i}}(\mathbf{y})))$.
Another way of stating this is that we assume for $u$ a piecewise
decomposition 
\begin{equation}
u=\begin{cases}
u^{0}\,, & x\in\mathscr{D}^{0}\,,\\
 & \vdots\\
u^{M}\,, & x\in\mathscr{D}^{M}\,,
\end{cases}
\end{equation}
where the $\mathscr{D}^{i}$'s are disjoint subsets of $\mathscr{I}$
between each ``particle location'', i.e.  
\begin{equation}
\mathscr{I}={\rm supp}\left(S\right)\cup\left(\bigcup_{i=0}^{M}\mathscr{D}^{i}\right)\,,
\end{equation}
where
\begin{equation}
\mathscr{D}^{i}=\begin{cases}
\left\{ x\in\mathscr{I}|x<x_{p_{1}}\right\} \,, & i=1\,,\\
\left\{ x\in\mathscr{I}|x_{p_{i}}<x<x_{p_{i+1}}\right\} \,, & 1\leq i\leq M-1\,,\\
\left\{ x\in\mathscr{I}|x_{p_{M}}<x\right\} \,, & i=M\,.
\end{cases}
\end{equation}

The general strategy, then, is to insert (\ref{eq:pwp-decomposition_general})
into (\ref{eq:setup-pde_general_2}), and to obtain a set of equations
by matching (regular function) terms multiplying the same derivative
order of the Heaviside distributions. Explicitly, using the Leibniz rule, we get
\begin{equation}
\mathcal{L}u=\sum_{i=0}^{M}\sum_{|\alpha|\leq m}\sum_{|\beta|\leq|\alpha|}\binom{\alpha}{\beta}\xi^{\alpha}\left(D^{\alpha-\beta}u^{i}\right)\left(D^{\beta}\Theta^{i}\right)=\sum_{i=1}^{M}\sum_{j=0}^{m-1}f^{ij}\delta_{\left(p_{i}\right)}^{(j)}\,.\label{eq:pwp-pde_decomposed}
\end{equation}

At zeroth order in derivatives of the Heaviside functions, i.e. the sum of
all $|\beta|=0$ terms in the LHS above, we will always simply obtain—in
the absence of any Heaviside functions on the RHS—a set of $M+1$ homogeneous
equations, which constitute simply the original equation on each disjoint
subset of $\mathscr{I}$ but \emph{with no source}: 
\begin{equation}
\sum_{i=0}^{M}\Bigg(\sum_{|\alpha|\leq m}\xi^{\alpha}D^{\alpha}u^{i}\Bigg)\Theta^{i}=0\Leftrightarrow\mathcal{L}u^{i}=0\enskip\mbox{in}\enskip\mathscr{D}^{i}\times\mathscr{V},\,\forall i\,.\label{eq:pwp-homogeneous_equations}
\end{equation}

At first order and higher in the Heaviside derivatives (thus, zeroth
order and higher in delta function derivatives), i.e. the sum of all
$|\beta|\neq0$ terms in the LHS of (\ref{eq:pwp-pde_decomposed}),
we have terms of the form 
\begin{align}
D^{\beta}\Theta^{i}=\, & \partial_{x}^{\beta_{0}}\partial_{y_{1}}^{\beta_{1}}\cdots\partial_{y_{n-1}}^{\beta_{n-1}}\begin{cases}
\Theta_{\left(p_{1}\right)}^{-}\,, & i=0\,,\\
\Theta_{\left(p_{i}\right)}^{+}-\Theta_{\left(p_{i+1}\right)}^{+}\,, & 1\leq i\leq M-1\,,\\
\Theta_{\left(p_{M}\right)}^{+}\,, & i=M\,,
\end{cases}\\
=\, & \sum_{j=0}^{|\beta|-1}\begin{cases}
F^{0j}\delta_{\left(p_{1}\right)}^{\left(j\right)}\,, & i=0\,,\\
F^{ij}\delta_{\left(p_{i}\right)}^{\left(j\right)}+G^{ij}\delta_{\left(p_{i+1}\right)}^{\left(j\right)}\,, & 1\leq i\leq M-1\,,\\
F^{Mj}\delta_{\left(p_{M}\right)}^{\left(j\right)}\,, & i=M\,,
\end{cases}\label{eq:pwp-D_beta_Theta2}
\end{align}
for some $\mathbf{y}$-dependent functions $F^{ij}:\mathscr{V}\rightarrow\mathbb{R}$
and $G^{ij}:\mathscr{V}\rightarrow\mathbb{R}$ which arise from the
implicit differentiation (e.g., Eqns. (\ref{eq:setip-partial_x_Theta})-(\ref{eq:setup-partial_y_Theta})),
and the precise form of which does not concern us for the present
purposes. Plugging (\ref{eq:pwp-D_beta_Theta2}) into (\ref{eq:pwp-pde_decomposed})
and manipulating the sums, we get 
\begin{equation}
\sum_{i=1}^{M}\sum_{|\alpha|\leq m}\sum_{0<|\beta|\leq|\alpha|}\sum_{j=0}^{|\beta|-1}\Phi^{\alpha,\beta,ij}\delta_{\left(p_{i}\right)}^{\left(j\right)}=\sum_{i=1}^{M}\sum_{j=0}^{m-1}f^{ij}\delta_{\left(p_{i}\right)}^{\left(j\right)}\,.\label{eq:pwp-pde_beta_nonzero}
\end{equation}
where for convenience we have defined
\begin{equation}
\Phi^{\alpha,\beta,ij}\left(\mathbf{x}\right)=\binom{\alpha}{\beta}\xi^{\alpha}\left(\mathbf{x}\right)\left(F^{ij}\left(\mathbf{y}\right)D^{\alpha-\beta}u^{i}\left(\mathbf{x}\right)+H^{ij}\left(\mathbf{y}\right)D^{\alpha-\beta}u^{i-1}\left(\mathbf{x}\right)\right)\,,
\end{equation}
for some $\mathbf{y}$-dependent functions $H^{ij}:\mathscr{V}\rightarrow\mathbb{R}$
(related to $F^{ij}$ and $G^{ij}$, and the precise form of which
is also unimportant). At this point, one must be careful: \emph{before}
drawing conclusions regarding the equality of terms (the coefficients
of the delta function derivatives) in (\ref{eq:pwp-pde_beta_nonzero}),
one should apply the identity (\ref{eq:setup-distributions_identity}).
Doing this, one obtains: 
\begin{equation}
\sum_{\alpha,\beta,i,j}\sum_{k=0}^{j}\left(-1\right)^{j+k}\binom{j}{k}\left(\partial_{x}^{j-k}\Phi^{\alpha,\beta,ij}\right)_{p_{i}}\delta_{\left(p_{i}\right)}^{\left(k\right)}=\sum_{i,j}\sum_{k=0}^{j}\left(-1\right)^{j+k}\binom{j}{k}\left(\partial_{x}^{j-k}f^{ij}\right)_{p_{i}}\delta_{\left(p_{i}\right)}^{\left(k\right)}\,,\label{eq:pwp-pde_with_identity}
\end{equation}
with the omitted summation limits as before. Thus, we see that on
the LHS, we have terms involving 
\begin{align}
\partial_{x}^{j-k}\Phi^{\alpha,\beta,ij}=\, & \partial_{x}^{j-k}\left\{ \binom{\alpha}{\beta}\xi^{\alpha}\left(F^{ij}D^{\alpha-\beta}u^{i}+H^{ij}D^{\alpha-\beta}u^{i-1}\right)\right\} \\
=\, & \binom{\alpha}{\beta}\left\{ F^{ij}\partial_{x}^{j-k}\left(\xi^{\alpha}D^{\alpha-\beta}u^{i}\right)+H^{ij}\partial_{x}^{j-k}\left(\xi^{\alpha}D^{\alpha-\beta}u^{i-1}\right)\right\} \\
=\, & \binom{\alpha}{\beta} \sum_{l=0}^{j-k}\binom{j-k}{l}\left(\partial_{x}^{j-k-l}\xi^{\alpha}\right)\left[F^{ij}\left(\partial_{x}^{l}D^{\alpha-\beta}u^{i}\right)+H^{ij}\left(\partial_{x}^{l}D^{\alpha-\beta}u^{i-1}\right)\right] \,.\label{eq:pwp-partial_x_Phi}
\end{align}
Thus, defining the $\mathbf{y}$-dependent functions 
\begin{align}
\Psi^{\alpha,\beta,ijkl}=\, & \left(-1\right)^{j+k}\binom{j}{k}\binom{j-k}{l}\binom{\alpha}{\beta}\left(\partial_{x}^{j-k-l}\xi^{\alpha}\right)_{p_{i}}\,,\\
\psi^{ijk}=\, & \left(-1\right)^{j+k}\binom{j}{k}\left(\partial_{x}^{j-k}f^{ij}\right)_{p_{i}}\,,
\end{align}
we can use (\ref{eq:pwp-partial_x_Phi}) to write (\ref{eq:pwp-pde_with_identity})
in the form: 
\begin{equation}
\sum_{\alpha,\beta,i,j,k,l}\Psi^{\alpha,\beta,ijkl}\left[F^{ij}\left(\partial_{x}^{l}D^{\alpha-\beta}u^{i}\right)_{p_{i}}+H^{ij}\left(\partial_{x}^{l}D^{\alpha-\beta}u^{i-1}\right)_{p_{i}}\right]\delta_{\left(p_{i}\right)}^{\left(k\right)}=\sum_{i,j,k}\psi^{ijk}\delta_{\left(p_{i}\right)}^{\left(k\right)}\,,\label{eq:pwp-pde_deltas_final}
\end{equation}
where the terms involving $u^{i}$ partials ``at the particle''
should be understood as the limit evaluated from the appropriate direction,
i.e. 
\begin{align}
\left(D^{\gamma}u^{i}\left(\mathbf{x}\right)\right)_{p_{i}}=\, & \lim_{x\rightarrow x_{p_{i}}^{+}}D^{\gamma}u^{i}\left(x,\mathbf{y}\right)\,,\\
\left(D^{\gamma}u^{i-1}\left(\mathbf{x}\right)\right)_{p_{i}}=\, & \lim_{x\rightarrow x_{p_{i}}^{-}}D^{\gamma}u^{i}\left(x,\mathbf{y}\right)\,.
\end{align}
Having obtained (\ref{eq:pwp-pde_deltas_final}), we can finally match
the coefficients of each $\delta_{\left(p_{i}\right)}^{\left(k\right)}$
to obtain the JCs with which the homogeneous equations (\ref{eq:pwp-homogeneous_equations})
must be supplemented.

Let us now extend this method to problems where the maximum order
of delta function derivatives in the source equals or exceeds the
order of the PDE, i.e. $K\geq m$, a case not previously required---and hence not yet considered---in any of the past PwP work on the self-force. To do this, we just add to our
ansatz the second term on the RHS of (\ref{eq:pwp-decomposition}),
which for convenience we denote $u^{\delta}$; that is: 
\begin{equation}
u=\sum_{i=0}^{M}u^{i}\Theta^{i}+u^{\delta}\,,\quad u^{\delta}=\sum_{i=1}^{M}\sum_{j=0}^{K-m}h^{ij}\delta_{\left(p_{i}\right)}^{\left(j\right)}\,,\label{eq:pwp-decomposition_general_with_delta}
\end{equation}
with $h^{ij}(\mathbf{y})$ to be solved for. Inserting (\ref{eq:pwp-decomposition_general_with_delta})
into (\ref{eq:setup-pde_general_2}) we get, on the LHS of the PDE,
the homogeneous problems (at zeroth order) as before, then the LHS
of (\ref{eq:pwp-pde_deltas_final}) due again to the sum of Heaviside functions
term in (\ref{eq:pwp-decomposition_general_with_delta}), plus the
following due to the sum of delta function derivatives: 
\begin{align}
\mathcal{L}u^{\delta}=\, & \sum_{|\alpha|\leq m}\xi^{\alpha}D^{\alpha}\sum_{i=1}^{M}\sum_{j=0}^{K-m}h^{ij}\delta_{\left(p_{i}\right)}^{\left(j\right)}\\
=\, & \sum_{\alpha,i,j}\sum_{|\beta|\leq|\alpha|}\binom{\alpha}{\beta}\xi^{\alpha}\left(D^{\alpha-\beta}h^{ij}\right)\left(D^{\beta}\delta_{\left(p_{i}\right)}^{\left(j\right)}\right)\,,
\end{align}
using the Leibniz rule. Next, we employ the Faà di Bruno formula \cite{constantine_multivariate_1996} to
carry out the implicit differentiation of the delta function derivatives;
writing $(n-1)$ dimensional multi-indices on $\mathscr{V}$ (pertaining
only to the $\mathbf{y}$ variables) with tildes, e.g. $\tilde{\beta}=(\beta_{1},\ldots,\beta_{n-1})$,
we have the following: 
\begin{equation}
D^{\beta}\delta_{\left(p_{i}\right)}^{\left(j\right)}=D^{\tilde{\beta}}\delta_{\left(p_{i}\right)}^{(j+\beta_{0})}=\tilde{\beta}!\sum_{l=1}^{|\tilde{\beta}|}\delta_{\left(p_{i}\right)}^{(j+\beta_{0}+l)}\sum_{s=1}^{|\tilde{\beta}|}\sum_{\mathscr{P}_{s}(\tilde{\beta},l)}\prod_{k=1}^{s}\frac{\left(-D^{\tilde{\lambda}_{k}}x_{p_{i}}\right)^{q_{k}}}{q_{k}!\left(\tilde{\lambda}_{k}!\right)^{q_{k}}}\,,
\end{equation}
where $\mathscr{P}_{s}(\tilde{\beta},l)=\{(q_{1},\ldots,q_{s};\tilde{\lambda}_{1},\ldots,\tilde{\lambda}_{s}):q_{k}>0,0\prec\tilde{\lambda}_{1}\prec\cdots\prec\tilde{\lambda}_{s},\sum_{k=1}^{s}q_{k}=l$
and $\sum_{k=1}^{s}q_{k}\tilde{\lambda}_{k}=\tilde{\beta}\}$. Therefore,
with all the summation limits the same as above, we get 
\begin{equation}
\mathcal{L}u^{\delta}=\sum_{\alpha,\beta,i,j,l,s}\sum_{\mathscr{P}_{s}(\tilde{\beta},l)}\binom{\alpha}{\beta}\tilde{\beta}!\xi^{\alpha}\left(D^{\alpha-\beta}h^{ij}\right)\delta_{\left(p_{i}\right)}^{(j+\beta_{0}+l)}\prod_{k=1}^{s}\frac{\left(-D^{\tilde{\lambda}_{k}}x_{p_{i}}\right)^{q_{k}}}{q_{k}!\left(\tilde{\lambda}_{k}!\right)^{q_{k}}}\,.
\end{equation}
Finally, we use the distributional identity (\ref{eq:setup-distributions_identity})
to obtain 
\begin{align}
\mathcal{L}u^{\delta}\equiv\, & \sum_{\alpha,\beta,i,j,l,s}\sum_{\mathscr{P}_{s}(\tilde{\beta},l)}\binom{\alpha}{\beta}\tilde{\beta}!\left(\left(D^{\alpha-\beta}h^{ij}\right)\prod_{k=1}^{s}\frac{\left(-D^{\tilde{\lambda}_{k}}x_{p_{i}}\right)^{q_{k}}}{q_{k}!\left(\tilde{\lambda}_{k}!\right)^{q_{k}}}\right)_{p_{i}}\nonumber \\
 & \times\left(-1\right)^{j+\beta_{0}+l}\sum_{r=0}^{j+\beta_{0}+l}\left(-1\right)^{r}\binom{j+\beta_{0}+l}{r}\left(\partial_{x}^{j+\beta_{0}+l-r}\xi^{\alpha}\right)_{p_{i}}\delta_{\left(p_{i}\right)}^{\left(r\right)}\,,
\end{align}
with which the higher order delta function derivatives on the RHS
of (\ref{eq:setup-pde_general_2}) can be matched.

\subsection{Limitations of the method}

Let us now discuss more amply the potential issues one is liable to
encounter in any attempt to extend the PwP method further beyond the
setup we have described so far.

Firstly, we stress once more that the method is applicable only to
\emph{linear} PDEs. As pointed out in the introduction, this is simply
an inherent limitation of the classic theory of distributions. In
particular, there it has long been proved \cite{schwartz_sur_1954} (see also the discussion in Ref. \cite{bottazzi_grid_2017}) that there does not exist
a differential algebra $(A,+,\otimes,\delta)$ wherein the real distributions
can be embedded, and: (i) $\otimes$ extends the product over $C^{0}(\mathbb{R})$;
(ii) $\delta:A\rightarrow A$ extends the distributional derivative;
(iii) $\forall u,v\in A$, the product rule $\delta(u\otimes v)=(\delta u)\otimes v+u\otimes(\delta v)$
holds. Attempts have been made to overcome this and create a sensible
nonlinear theory of distributions by defining and working with more
general objects dubbed ``generalized functions'' \cite{colombeau_nonlinear_2013}. Nonetheless, these
have their own drawbacks (e.g. they sacrifice coherence between the
product over $C^{0}(\mathbb{R})$ and that of the differential
algebra), and different formulations are actively being investigated
by mathematicians \cite{benci_ultrafunctions_2013,bottazzi_grid_2017}. A PwP method for nonlinear problems in the context
of these formulations could be an interesting line of inquiry for
future work.

Secondly, as we have seen, the PwP method as developed here is guaranteed
to work only for those (linear) PDEs the source $S$ of which is a distribution
not on the entire problem domain $\mathscr{U}$, but only on a one-dimensional
subspace $\mathscr{I}$ of that domain. One may sensibly wonder whether
this situation can be improved, i.e. whether a similar procedure could
succeed in tackling equations with sources involving (derivatives
of) delta functions in \emph{multiple} variables—yet, one may also
immediately realize that such an attempted extension quickly leads
to significant complications and potentially impassable problems.
Let us suppose that the source contains (derivatives of) delta functions
in $\bar{n}>1$ variables. We still define $\mathscr{I}$ such that
${\rm supp}(S)\subset\mathscr{I}$, so now we have $\mathscr{I}\subseteq\mathbb{R}^{\bar{n}}$,
and let us adapt the rest of our notation accordingly so that an arbitrary
function on $\mathscr{U}$ is
\begin{align}
f:\mathscr{U}\,=\,\mathscr{I}\times\mathscr{V}\,\,\subseteq\,\,\mathbb{R}^{\bar{n}}\times\mathbb{R}^{n-\bar{n}}\,\,=\,\,\mathbb{R}^{n} & \rightarrow\mathbb{R}\nonumber \\
\mathbf{x}=\left(\bar{{\bf x}},\mathbf{y}\right)=\left(\bar{x}_{1},...,\bar{x}_{\bar{n}},y_{1},...,y_{n-\bar{n}}\right) & \mapsto f\left(\mathbf{x}\right)\,.
\end{align}
We also adapt the multi-index notation to $\alpha=(\bar{\alpha}_{1},\bar{\alpha}_{2},\ldots,\bar{\alpha}_{\bar{n}},\alpha_{1},\alpha_{2},\ldots,\alpha_{n-\bar{n}})$.
We can still write the most general linear partial differential operator,
just as we did earlier, as $\mathcal{L}=\sum_{|\alpha|\leq m}\xi^{\alpha}(\mathbf{x})D^{\alpha}$
where now $D^{\alpha}=\partial^{|\alpha|}/\partial\bar{x}_{1}^{\bar{\alpha}_{1}}\cdots\partial\bar{x}_{\bar{n}}^{\bar{\alpha}_{\bar{n}}}\partial y_{1}^{\alpha_{1}}\cdots\partial y_{n-\bar{n}}^{\alpha_{n-\bar{n}}}$.
Moreover, in general, we use the barred boldface notation $\bar{{\bf v}}$
for any vector in $\mathscr{I}$, $\bar{{\bf v}}=(\bar{v}_{1},\ldots,\bar{v}_{\bar{n}})\in\mathscr{I}\subseteq\mathbb{R}^{\bar{n}}$.

One may first ask whether a PwP-type method could be used to handle
``point'' sources in $\mathscr{I}\subseteq\mathbb{R}^{\bar{n}}$.
In other words, can we find a decomposition of $u$ which could be
useful for a problem of the form 
\begin{equation}
\mathcal{L}u\left(\mathbf{x}\right)=f\left(\mathbf{x}\right)\delta\left(\bar{{\bf x}}-\bar{{\bf x}}_{p}\left(\mathbf{y}\right)\right)+\bar{{\bf g}}\left(\mathbf{x}\right)\cdot\bar{\boldsymbol{\nabla}}\delta\left(\bar{{\bf x}}-\bar{{\bf x}}_{p}\left(\mathbf{y}\right)\right)+\cdots\,,\label{eq:pwp-multivariable_source_Lu}
\end{equation}
(assuming for simplicity a \emph{single} point source at $\bar{{\bf x}}_{p}\in\mathscr{I}$)
with $\bar{\boldsymbol{\nabla}}=\partial/\partial\bar{{\bf x}}$ and
given functions $f:\mathscr{U\rightarrow\mathbb{R}}$, $\bar{{\bf g}}:\mathscr{U}^{\bar{n}}\rightarrow\mathbb{R}$
etc.? Intuitively, in order to match the delta function (derivatives)
on the RHS, we might expect $u$ to contain the $\bar{n}$-dimensional
Heaviside function $\Theta:\mathcal{D}(\mathscr{I})\rightarrow\mathbb{R}$.
Thus, in the same vein as (\ref{eq:pwp-decomposition_general}), a
possible attempt (for $K<m$) might be to try a splitting such as
\begin{equation}
u\left(\mathbf{x}\right)=\sum_{\bar{\boldsymbol{\sigma}}=\mathrm{\Pi}^{\bar{n}}(\pm)}u^{\bar{\boldsymbol{\sigma}}}\left(\mathbf{x}\right)\Theta\left(\bar{\boldsymbol{\sigma}}\odot\left(\bar{{\bf x}}-\bar{{\bf x}}_{p}\left(\mathbf{y}\right)\right)\right)\,,\label{eq:pwp-multivariable_source_u}
\end{equation}
where $\mathrm{\Pi}$ is here the Cartesian product and $\odot$ the entrywise
product; \emph{but} whether or not this will work depends completely
upon the detailed form of $\mathcal{L}$. For example, the procedure
\textsl{might} work in the case where $\mathcal{L}$ contains a nonvanishing
$D^{(1,1,\ldots,1,\alpha_{1},...,\alpha_{n-\bar{n}})}$ term, so as
to produce a $\delta(\bar{{\bf x}}-\bar{{\bf x}}_{p})$ term upon
its action on $u$ (in the form (\ref{eq:pwp-multivariable_source_u})),
needed to match the $f(\mathbf{x})$ term on the RHS of (\ref{eq:pwp-multivariable_source_Lu}).
However, this still does not guarantee that \textsl{all} the distributional
terms can in the end be appropriately matched, and so in general,
one should \emph{not} expect that such an approach in these sorts
of problems will yield a workable strategy.

To render the above discussion a little less abstract, let us illustrate
what we mean by way of a very simple example. Consider a two-dimensional
Poisson equation on $\mathscr{U}=\{(x,y)\}\subseteq\mathbb{R}^{2}$:
$(\partial_{x}^{2}+\partial_{y}^{2})u=\delta_{2}(x,y)$, where the
RHS is the two-dimensional delta function supported at the origin.
An attempt to solve this via our method would begin by decomposing
the solution into a form $u=\sum_{j}u^{j}\Theta^{j}$, for some suitably-defined
Heaviside functions $\Theta^{j}$--- supported, for example,
on positive/negative half-planes in each of the two coordinates, or
perhaps on each quadrant of $\mathbb{R}^{2}$. However, the RHS of
this problem is, by definition, $\delta_{2}(x,y)=\delta(x)\delta(y)=(\partial_{x}\Theta^{+}(x))(\partial_{y}\Theta^{+}(y))$,
and there is no way to get such a term from the operator $\mathcal{L}=\partial_{x}^{2}+\partial_{y}^{2}$
acting on any linear combination of Heaviside functions. The unconvinced
reader is invited to try a few attempts for themselves, and the difficulties
with this will quickly become apparent.

That said, \emph{one} case in which a PwP-type procedure could work
is when the source contains (one-dimensional) ``string''-like singularities
(instead of $\bar{n}$-dimensional ``point''-like ones) in each of
the $\bar{{\bf x}}$ variables---in other words, when our problem
is of the form 
\begin{equation}
\mathcal{L}u\left(\mathbf{x}\right)=\sum_{a=1}^{\bar{n}}f^{a}\left(\mathbf{x}\right)\delta\left(\bar{x}_{a}-\bar{x}_{a,p}\left(\mathbf{y}\right)\right)+\sum_{a=1}^{\bar{n}}g^{a}\left(\mathbf{x}\right)\delta'\left(\bar{x}_{a}-\bar{x}_{a,p}\left(\mathbf{y}\right)\right)+\cdots\,,
\end{equation}
with $f^{a}:\mathscr{U}\rightarrow\mathbb{R}$, $g^{a}:\mathscr{U}\rightarrow\mathbb{R}$
etc. Then, a decomposition of $u$ which can be tried in such situations
(for $K<m$) is 
\begin{equation}
u\left(\mathbf{x}\right)=\sum_{a=1}^{\bar{n}}\sum_{\sigma_{a}=\pm}u^{a,\sigma_{a}}\left(\mathbf{x}\right)\Theta\left(\sigma_{a}\left(\bar{x}_{a}-\bar{x}_{a,p}\left(\mathbf{y}\right)\right)\right)\,.
\end{equation}

\section{First order hyperbolic PDEs}
\label{first-order-hyperbolic-pdes}

We now move on to applications of the PwP method, beginning with first
order hyperbolic equations. First we look at the standard advection
equation, and then a simple neural population model from neuroscience.
Finally, we consider another popular advection-type problem with a
distributional source---namely, the shallow water equations with
discontinuous bottom topography---and briefly explain why the PwP
method cannot be used in that case.

\subsection{Advection equation}

As a first very elementary illustration of our method, let us consider
the $(1+1)$-dimensional advection equation for $u(x,t)$ with a time
function singular point source at some $x=x_{*}$: 
\begin{equation}
\begin{cases}
\partial_{t}u+\partial_{x}u=g\left(t\right)\delta\left(x-x_{*}\right)\,, & x\in\mathscr{I}=\left[0,L\right],\enskip t>0\,,\\
u\left(x,0\right)=0\,, & u\left(0,t\right)=u\left(L,t\right)\,,
\end{cases}\label{eq:advection_pde}
\end{equation}
where we assume that the source time function $g(t)$ is smooth and
vanishes at $t=0$. On an unbounded spatial domain (i.e. $x\in\mathbb{R}$),
the exact solution of this problem is 
\begin{equation}
u_{\text{ex}}\left(x,t\right)=\left[\Theta\left(x-x_{*}\right)-\Theta\left(x-x_{*}-t\right)\right]g\left(t-\left(x-x_{*}\right)\right)\,,\label{eq:advection_exact_soln}
\end{equation}
i.e. the forward-translated source function in the right half
of the future light cone emanating from $x_{*}$. If we suppose that the source location
satisfies $x_{*}\in(0,L/2]$, then (\ref{eq:advection_exact_soln})
is also a solution of our problem (\ref{eq:advection_pde}) for $t\in[0,L-x_{*}]$.

This precise problem is treated in Ref. \cite{petersson_discretizing_2016} using a (polynomial) delta function approximation procedure,
with the following: $g(t)={\rm e}^{-(t-t_{0})^{2}/2}$, $t_{0}=8$,
$L=40$ and $x_{*}=10+\pi$. We numerically implement the exact same setup, but using our
PwP method: that is, we decompose $u=u^{-}\Theta^{-}+u^{+}\Theta^{+}$ where
$\Theta^{\pm}=\Theta(\pm(x-x_{*}))$. Inserting this into (\ref{eq:advection_pde}),
we get homogeneous PDEs $\partial_{t}u^{\pm}+\partial_{x}u^{\pm}=0$
to the left and right of the singularity, i.e. on $x\in\mathscr{D}^{-}=[0,x_{*}]$
and $x\in\mathscr{D}^{+}=[x_{*},L]$ respectively, along with a jump
in the solution $[u]_{*}=g(t)$ at the point of the source singularity.

The details of our numerical scheme are described in Appendix \ref{a-numerical-first-order-hyperbolic}. We also offer in Appendix \ref{a-psc} a brief description of the PSC methods and notation used therein.

\begin{figure}
\begin{center}
\includegraphics[scale=0.6]{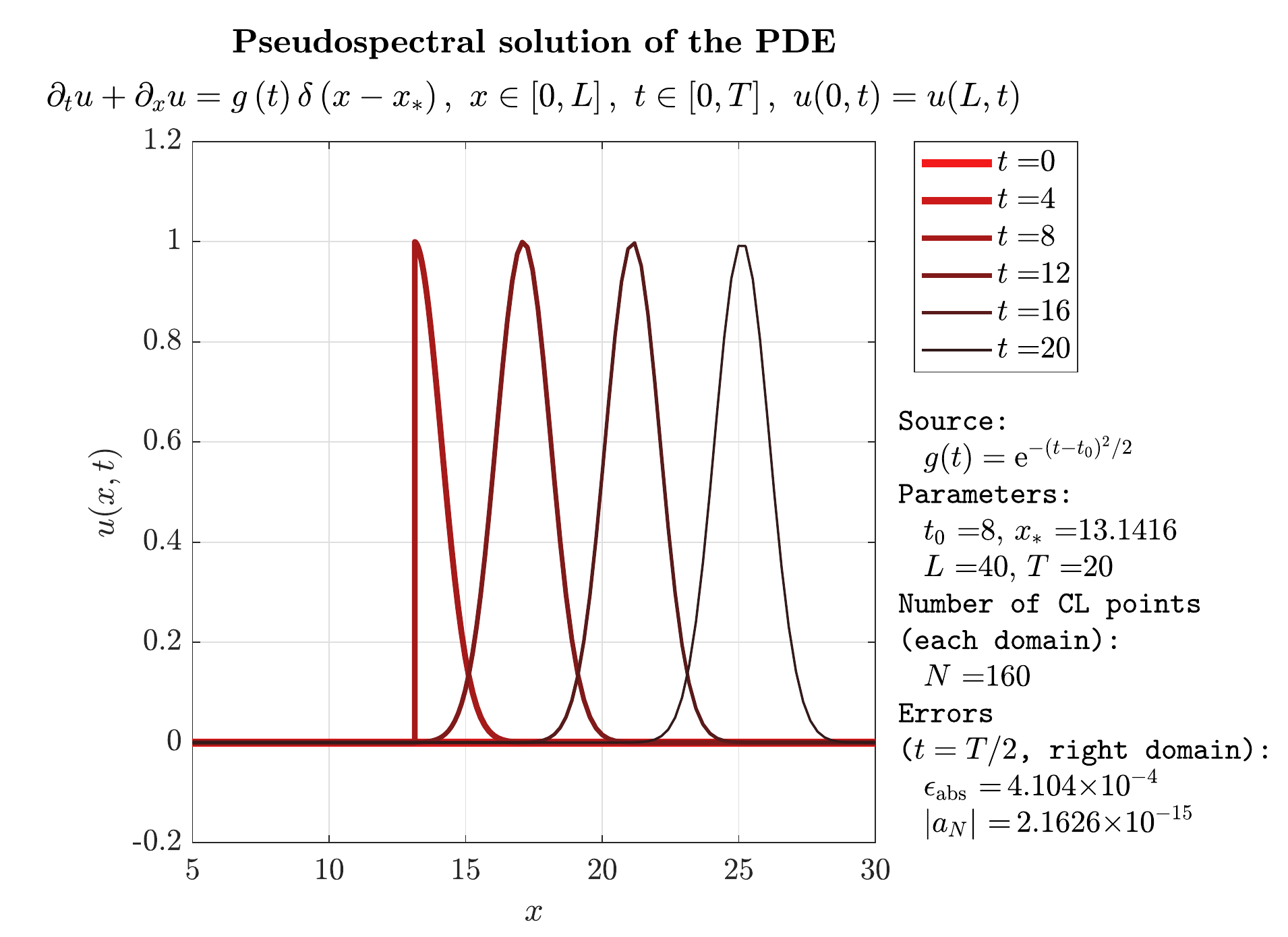}
\includegraphics[scale=0.6]{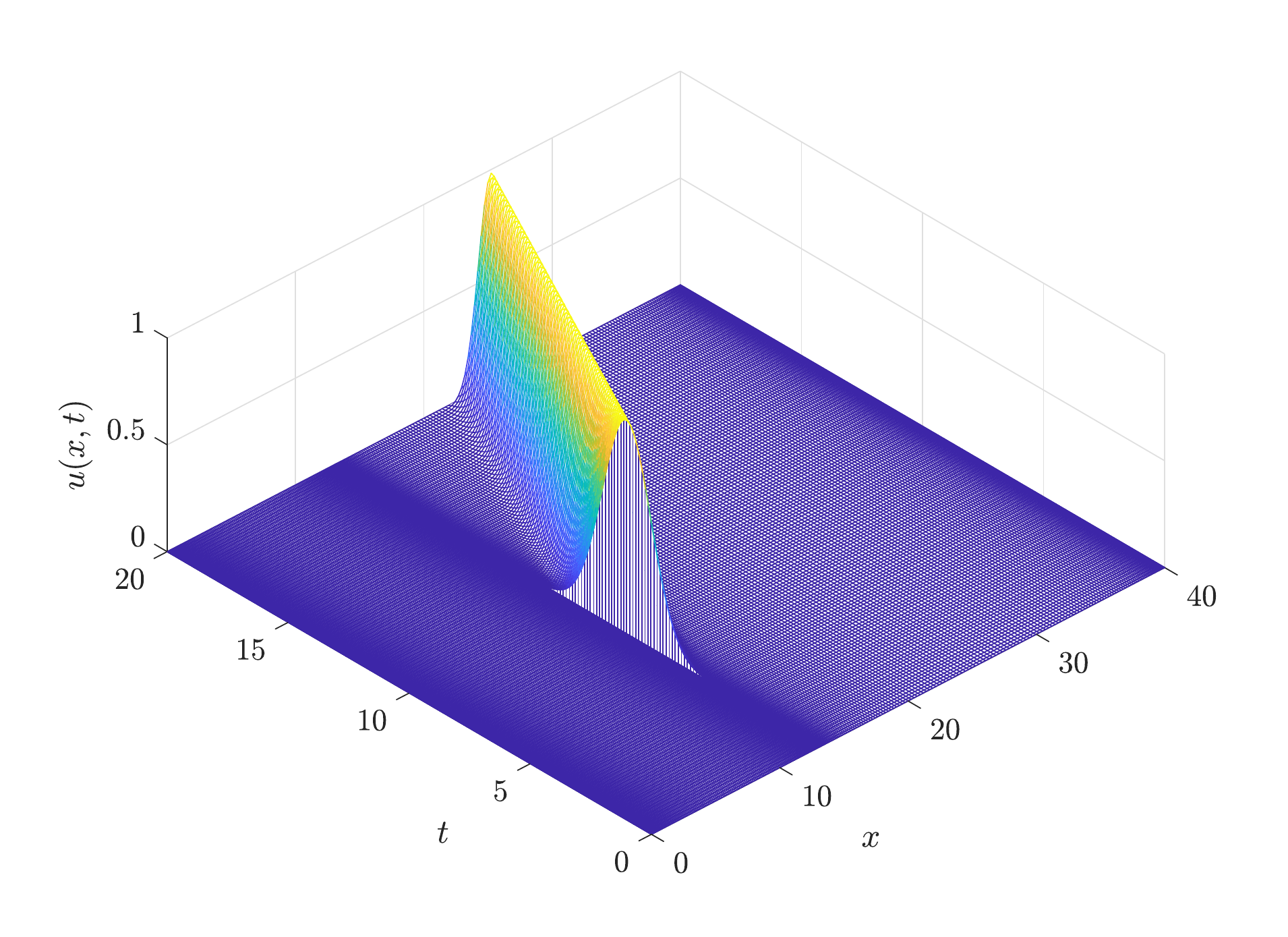}
\caption{Solution of the problem (\ref{eq:advection_pde}) with zero initial data.}\label{fig:pde_pwp_advection_soln}
\end{center}
\end{figure}

\begin{figure}
\begin{center}
\includegraphics[scale=0.6]{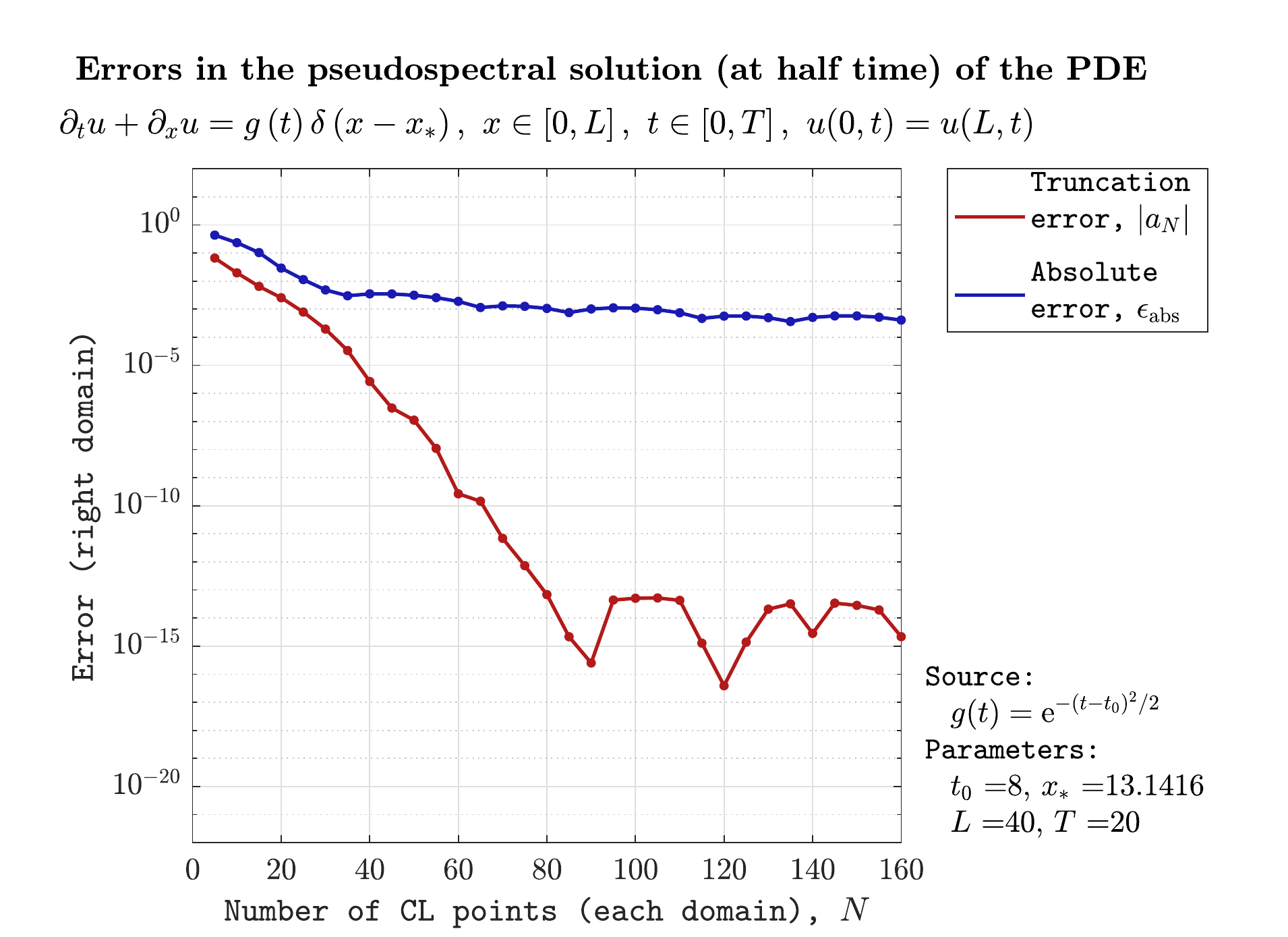}
\caption{Convergence of the numerical scheme for the problem (\ref{eq:advection_pde}).}\label{fig:pde_pwp_advection_error}
\end{center}
\end{figure}

The solution for zero initial data is displayed in Figure \ref{fig:pde_pwp_advection_soln}, and the numerical convergence in Figure \ref{fig:pde_pwp_advection_error}. For the latter, we plot---for the numerical solution $\bm{u}$
at $t=T/2$---both the absolute error (in the
$l^{2}$ norm on the CL grids, as in Ref. \cite{petersson_discretizing_2016}), $\epsilon_{{\text{abs}}}=||\bm{u}-\bm{u}_{{\text{ex}}}||_{2}$,
as well as the truncation error in the right CL domain $\mathscr{D}^{+}$
given simply the absolute value of the last spectral coefficient $a_{N}$
of $\bm{u}^{+}$. We see that the truncation error exhibits typical
(exponential) spectral convergence; the absolute error converges at
the same rate until $N\approx40$, after which it converges more slowly
because it becomes dominated by the $\mathcal{O}(\mathrm{\Delta} t)=\mathcal{O}(N^{-2})$
error in the finite difference time evolution scheme. Nevertheless, for the same number of grid points,
our procedure still yields a lower order of magnitude of the $l^{2}$ error as was obtained in Ref. \cite{petersson_discretizing_2016} with a \emph{sixth} order finite difference scheme (relying on a a source discretization with 6 moment conditions and 6 smoothness conditions); we present a simple comparison of these in the following table:

\begin{center}
\begin{tabular}{c|c|c}
$\epsilon_{\text{abs}}$ & $N=80$ & $N=160$\tabularnewline
\hline 
Ref. \cite{petersson_discretizing_2016} & $\mathcal{O}(10^{-2})$ & $\mathcal{O}(10^{-3})$\tabularnewline
\hline 
PwP method & $\mathcal{O}(10^{-3})$ & $\mathcal{O}(10^{-4})$\tabularnewline
\end{tabular}
\par\end{center}

\subsection{Advection-type equations in neuroscience}
Advection-type equations with distributional sources arise in practice,
for example, in the modeling of neural populations. In particular,
among the simplest of these are the so-called ``integrate-and-fire''
models. For some of the earlier work on such models from a neuroscience
perspective, see for example \cite{haskell_population_2001,casti_population_2002}
and references therein; for more recent work focusing on mathematical
aspects, see \cite{caceres_analysis_2011,caceres_blow-up_2016}. Their
aim is to describe the probability density $\rho\left(\bm{v},t\right)$
of neurons as a function of certain state variables $\bm{v}$ and
time $t$. Often the detailed construction of these models can be
quite involved and dependent on a large number of parameters, so to
simply illustrate the principle of our method we here consider the
simple case where the single state variable is the voltage $V$. Then,
generally speaking, the dynamics of $\rho(V,t)$ takes the form of
a Fokker-Planck-type equation on $V\in(-\infty,L]$ with a singular
source at some fixed $V=V_{*}<L$,
\begin{equation}
\partial_{t}\rho+\partial_{V}\left(f\left(V,N\left(t\right)\right)\rho\right)-\frac{\sigma^{2}}{2}\partial_{V}^{2}\rho=N\left(t\right)\delta\left(V-V_{*}\right)\,.\label{eq:fokker-planck}
\end{equation}
The source time function $N(t)$ must be such that conservation
of probability, i.e. $\partial_{t}\int{\rm d}V\rho=0$, is guaranteed
under homogeneous Dirichlet BCs.

As a simplification of this problem, let us suppose, as is sometimes done, that the diffusive
part (the second derivative term on the LHS) of (\ref{eq:fokker-planck})
is negligible. Moreover, in simple cases, the velocity function
$f$ in the advection term has the form $f=-V+{\rm constant}$, and
we just work with the constant set equal to $1$. We restrict ourselves
to a bounded domain for $V$ which for illustrative purposes we just
choose to be $\mathscr{I}=\left[0,L\right]$. Demanding homogeneous
Dirichlet BCs at the left boundary in conjunction with conservation
of probability fixes the source time function to be $N(t)=(1-L)\rho(L,t)$.
Thus, we are going to tackle the following problem: 
\begin{equation}
\begin{cases}
\partial_{t}\rho+\partial_{V}\left(\left(1-V\right)\rho\right)=\left(1-L\right)\rho\left(L,t\right)\delta\left(V-V_{*}\right)\,, & V\in\mathscr{I}=\left[0,L\right],t>0\,,\\
\rho\left(V,0\right)=\rho_{0}\left(V\right),\enskip\int_{\mathscr{I}}{\rm d}V\rho_{0}\left(V\right)=1\,, & \rho\left(0,t\right)=0\,.
\end{cases}\label{eq:neural_population}
\end{equation}

We now implement the PwP decomposition: $\rho=\rho^{-}\Theta^{-}+\rho^{+}\Theta^{+}$
with $\Theta^{\pm}=\Theta(\pm(V-V_{*}))$. Inserting this into the
PDE (\ref{eq:neural_population}), we get the homogeneous problems
$\partial_{t}\rho^{\pm}+\partial_{V}\left(\left(1-V\right)\rho^{\pm}\right)=0$
on $\mathscr{D}^{\pm}$, with $\mathscr{D}^{-}=[0,V_{*}]$ and $\mathscr{D}^{+}=[V_{*},L]$,
along with the JC $[\rho]_{*}=\frac{1-L}{1-V_{*}}\rho(L,t)$.

\begin{figure}
\begin{center}
\includegraphics[scale=0.6]{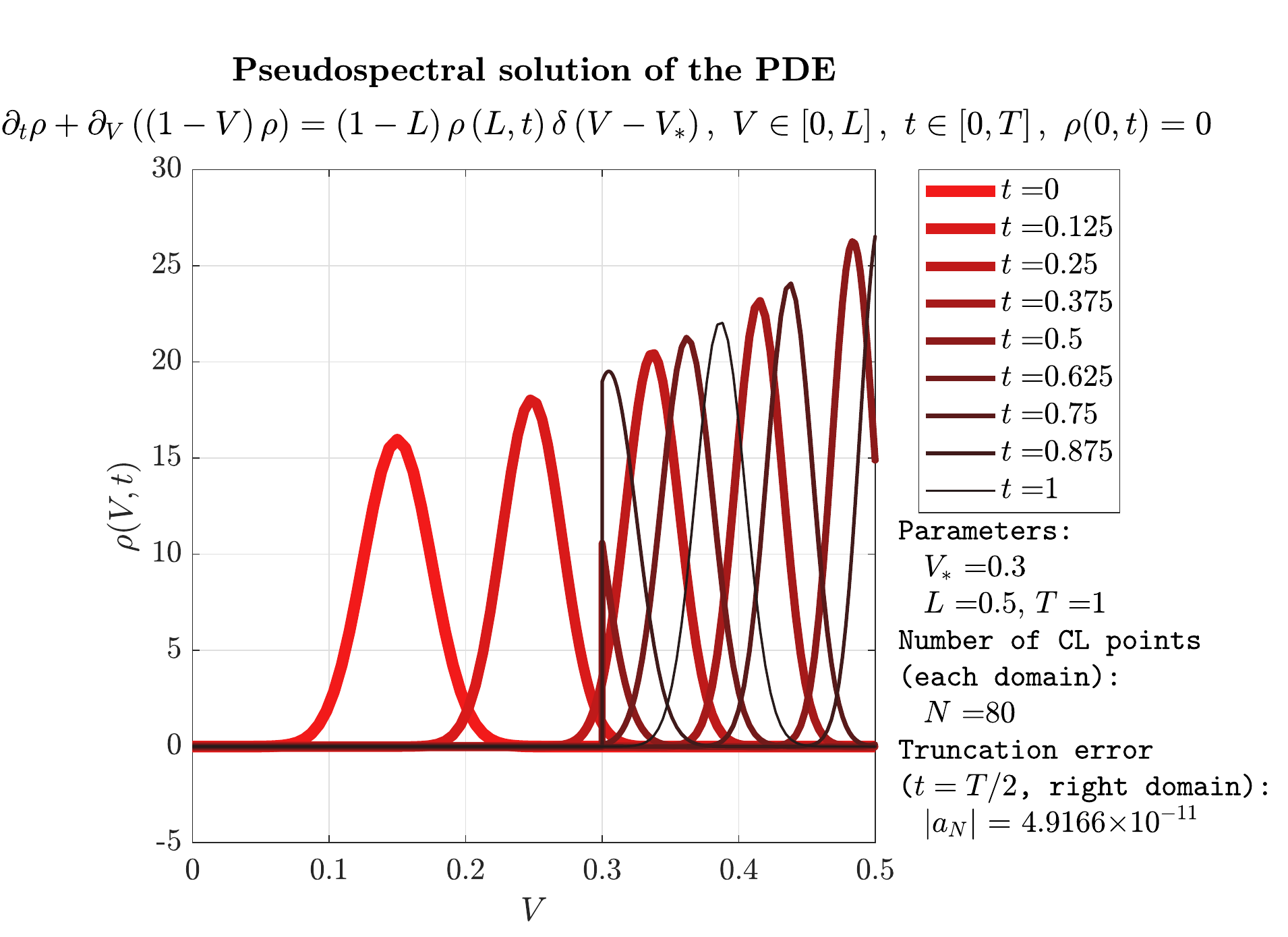}
\includegraphics[scale=0.6]{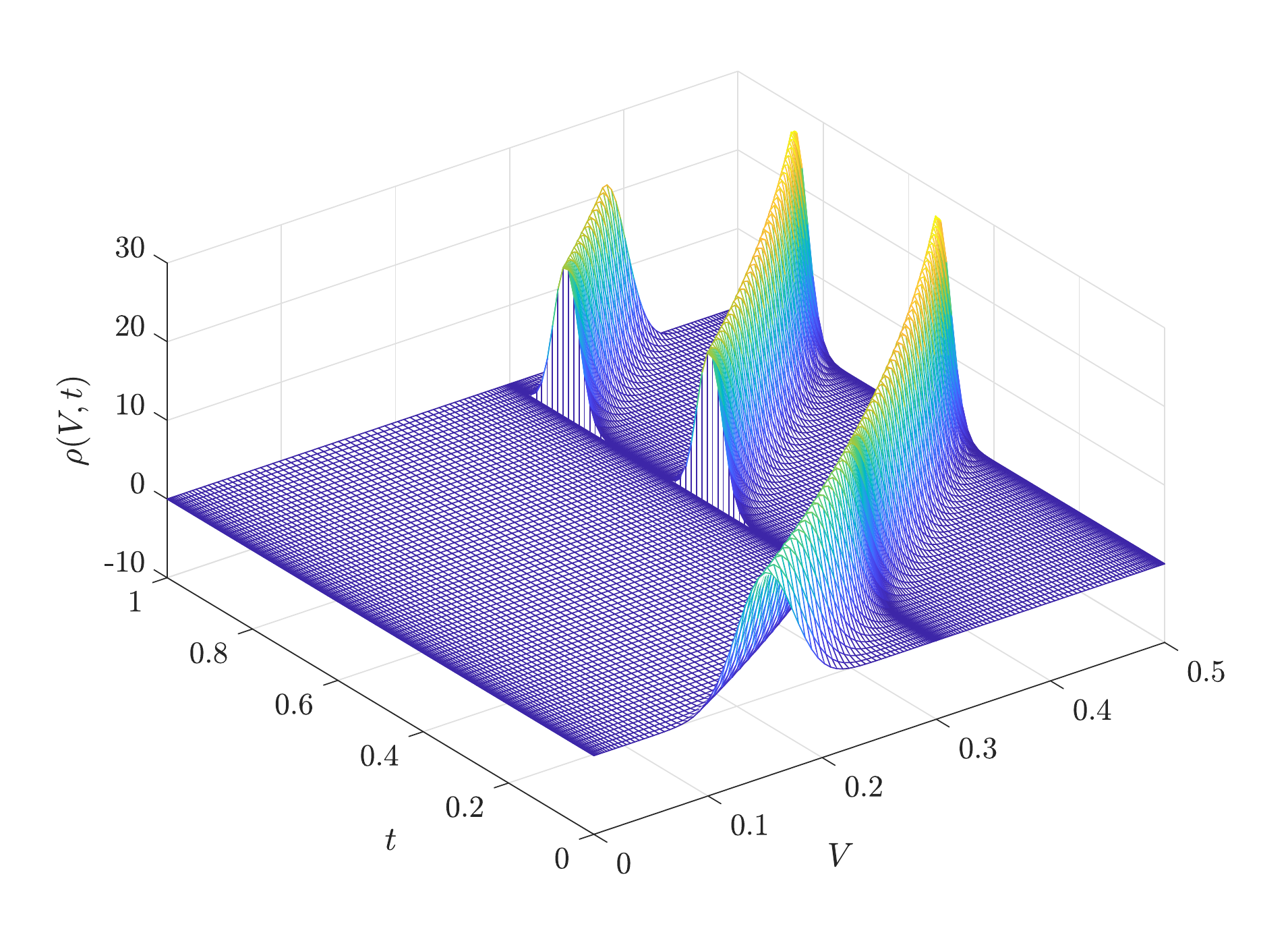}
\caption{Solution of (\ref{eq:neural_population}) with (normalized) Gaussian initial data centered at $V=0.3$.}\label{fig:pde_pwp_advection_neurosci_soln}
\end{center}
\end{figure}

\begin{figure}
\begin{center}
\includegraphics[scale=0.6]{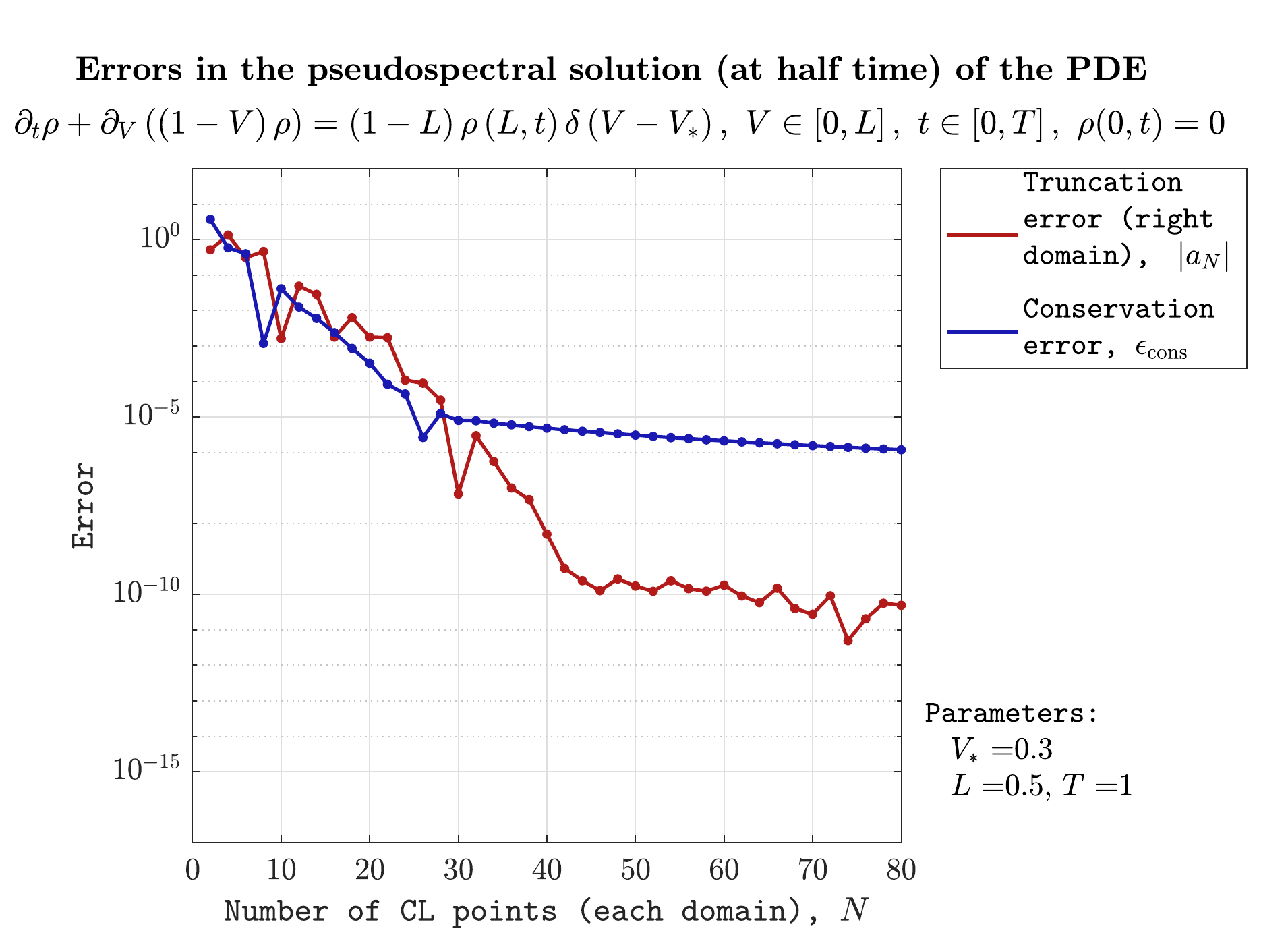}
\caption{Convergence of the numerical scheme for the problem (\ref{eq:neural_population}).}\label{fig:pde_pwp_advection_neurosci_error}
\end{center}
\end{figure}

An example solution for Gaussian initial data centered at $V=0.3$
is displayed in Figure \ref{fig:pde_pwp_advection_neurosci_soln},
and the numerical convergence in Figure \ref{fig:pde_pwp_advection_neurosci_error}.
In the latter, we plot---again for the numerical solution
$\bm{\rho}$ at the final time---the truncation error
as well as (in the absence of an exact solution) what we refer to as the conservation
error, $\epsilon_{{\text{cons}}}=|1-\int_{\mathscr{I}}{\rm d}V\rho(V,t)|$,
which simply measures how far we are from exact conservation of probability.
Both of these exhibit exponential convergence. The integral
in $\epsilon_{{\text{cons}}}$ is computed as a sum over both domains,
$\int_{\mathscr{I}}{\rm d}V\rho=\int_{\mathscr{D}^{-}}{\rm d}V\rho+\int_{\mathscr{D}^{+}}{\rm d}V\rho$,
and numerically performed on each using a standard pseudospectral
quadrature method (as in, e.g., Chapter 12 of Ref. \cite{trefethen_spectral_2001}).

This procedure can readily be complexified with the inclusion of a
diffusion term, and indeed we will shortly turn to purely diffusion (heat-type equation)
problems in the following section.

\subsection{Advection-type equations in other applications}

Another advection-type application in which one may be tempted to
try applying some form the PwP method is the shallow water equations.
Setting the gravitational acceleration to $1$, these read:
\begin{equation}
\partial_{t}\left[\begin{array}{c}
h\\
hu
\end{array}\right]+\partial_{x}\left[\begin{array}{c}
hu\\
hu^{2}+\tfrac{1}{2}h^{2}
\end{array}\right]=\left[\begin{array}{c}
0\\
-h\partial_{x}B
\end{array}\right]\,,\label{eq:shallow_water}
\end{equation}
where $B(x)$ is the elevation of the bottom topography, $h(x,t)$
is the fluid depth above the bottom and $u(x,t)$ is the velocity.
If the topography is discontinuous, i.e. if $B\notin C^{0}(\mathbb{R})$,
then the RHS of (\ref{eq:shallow_water}) will be distributional;
this can happen, e.g., if the bottom is a step (if $B=\Theta$, then
the RHS is $\icol{0\\-h\delta}$), a wall etc. However,
the problem with applying the PwP method here is that (\ref{eq:shallow_water})
is nonlinear, and so one encounters precisely the sorts of issues
detailed at the end of the preceding section. Indeed, explicit numerical
solutions that have been obtained for (\ref{eq:shallow_water}) in
the literature \cite{zhou_numerical_2002,bernstein_central-upwind_2016} qualitatively indicate that a PwP-type decomposition as described
here would be inadequate (and, anyway, nonsensical mathematically)
for such problems.

\section{Parabolic PDEs}
\label{parabolic-pdes}

We begin by analyzing the standard heat equation and then move on
to an application in finance which includes two (time-dependent) singular
source terms. 

\subsection{Heat equation}

Let us consider now the $(1+1)$-dimensional heat equation for $u(x,t)$
with a constant point source at a time-dependent location $x=x_{p}(t)$,
with Dirichlet boundary conditions:
\begin{equation}
\begin{cases}
\partial_{t}u-\partial_{x}^{2}u=\lambda\delta\left(x-x_{p}\left(t\right)\right)\,, & x\in\mathscr{I}=\left[a,b\right],\enskip t>0\,,\\
u\left(x,0\right)=0\,, & u\left(a,t\right)=\alpha,\,u\left(b,t\right)=\beta\,.
\end{cases}\label{eq:heat_pde}
\end{equation}
In this case, we do not have the exact solution.

This problem is treated in \cite{tornberg_numerical_2004} using a delta function approximation procedure,
with the following setup: $\mathscr{I}=[0,1]$, $\alpha=0=\beta$ and $\lambda=10$; constant-valued and sinusoidal point source locations $x_{p}(t)$ are considered. We implement here the same, using our PwP method: we decompose $u=u^{-}\Theta^{-}+u^{+}\Theta^{+}$
where $\Theta^{\pm}=\Theta(\pm(x-x_{p}(t)))$. Inserting this into
(\ref{eq:heat_pde}), we get homogeneous PDEs $\partial_{t}u^{\pm}-\partial_{x}^{2}u^{\pm}=0$
to the left and right of the singularity, $x\in\mathscr{D}^{-}=[0,x_{p}(t)]$
and $x\in\mathscr{D}^{+}=[x_{p}(t),1]$ respectively; additionally,
we have the following JCs: $[u]_{p}=0$ and $[\partial_{x}u]_{p}=-\lambda$.

\begin{figure}
\begin{center}
\includegraphics[scale=0.6]{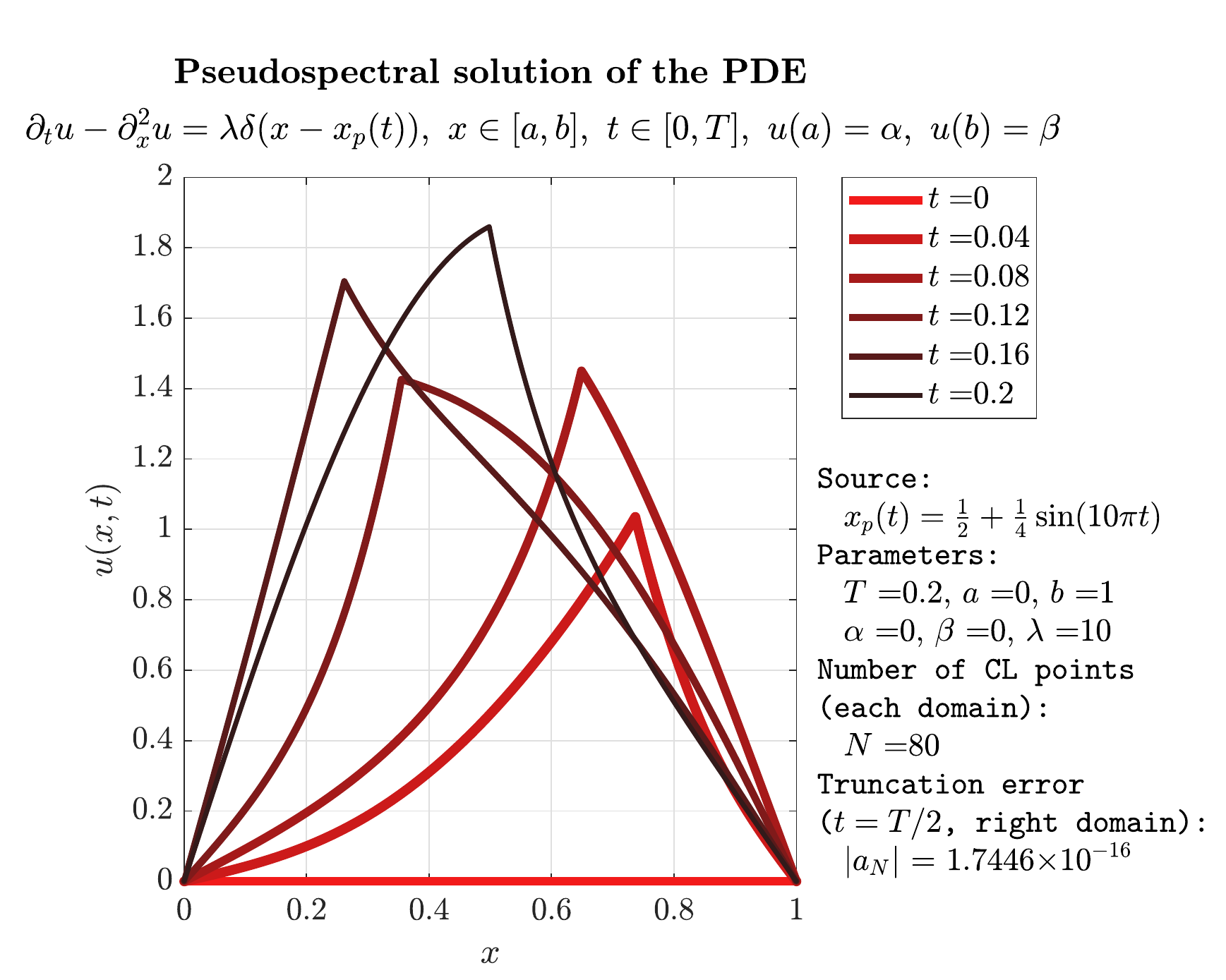}
\includegraphics[scale=0.6]{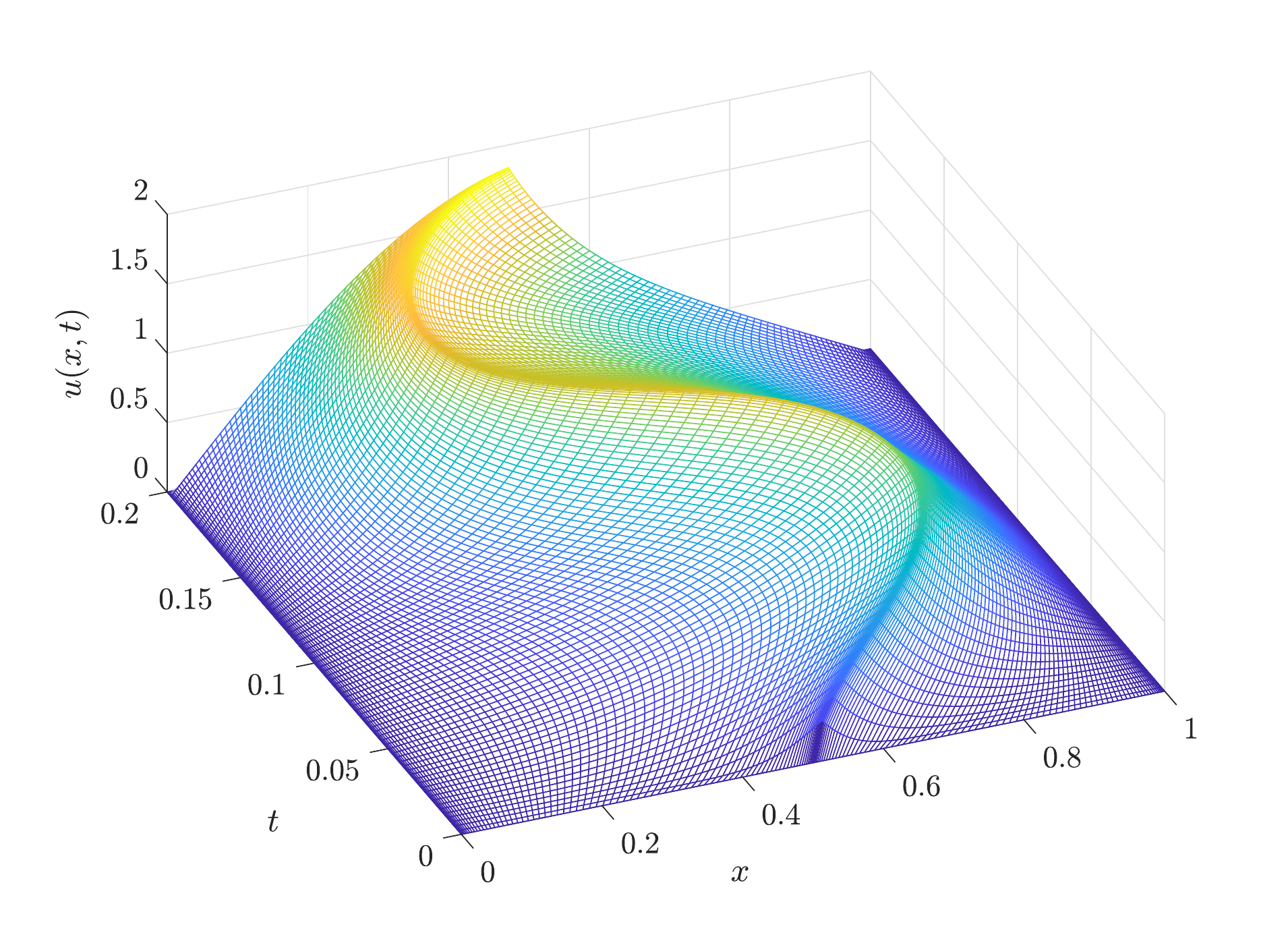}
\caption{Solution of the problem (\ref{eq:heat_pde}) with zero initial data.}\label{fig:pde_pwp_heat_soln}
\end{center}
\end{figure}

\begin{figure}
\begin{center}
\includegraphics[scale=0.6]{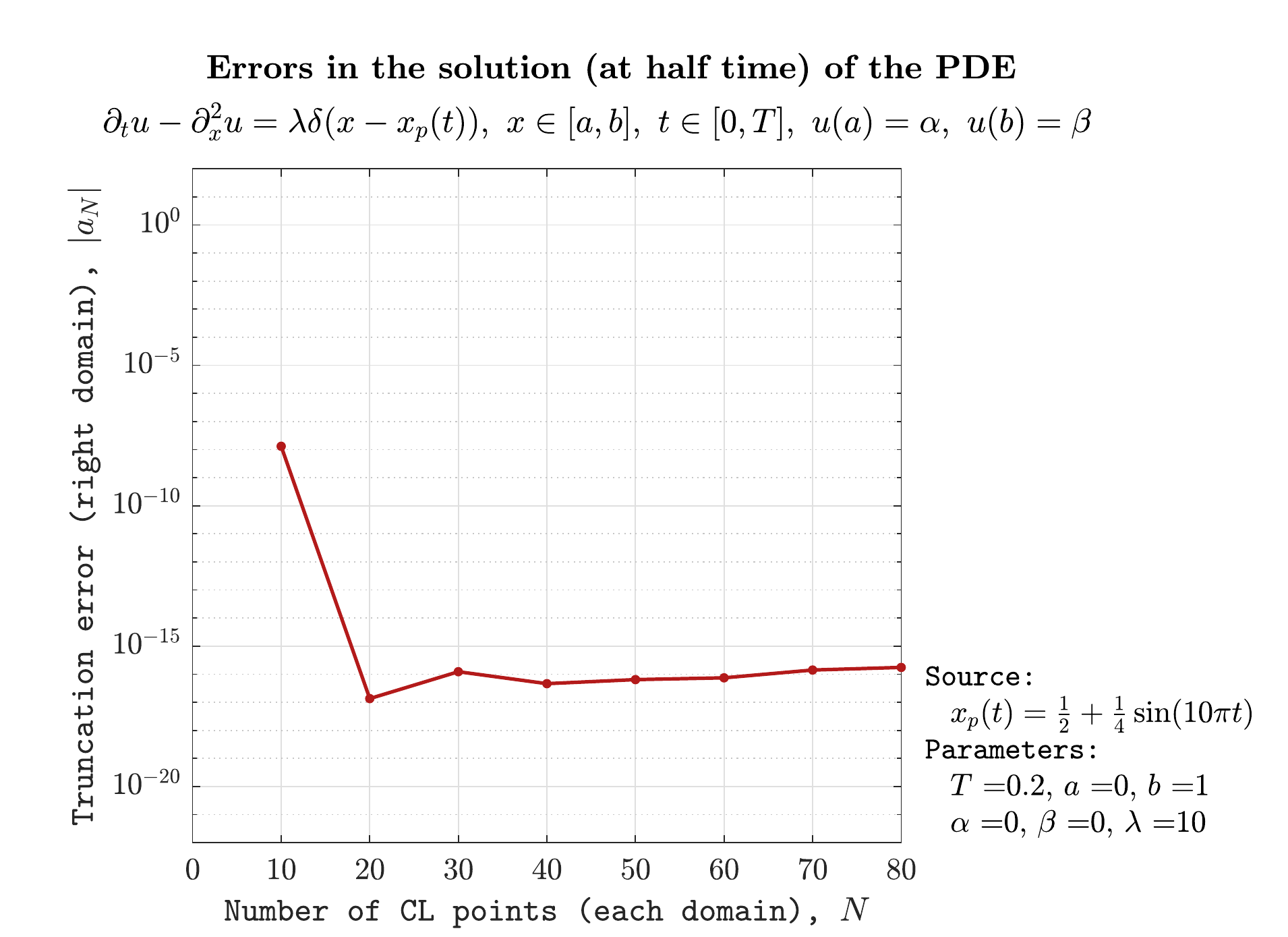}
\caption{Convergence of the numerical scheme for the problem (\ref{eq:heat_pde}).}\label{fig:pde_pwp_heat_error}
\end{center}
\end{figure}

The details of the numerical scheme are given in Appendix \ref{a-numerical-parabolic}, and results for zero initial data in Figures \ref{fig:pde_pwp_heat_soln} and \ref{fig:pde_pwp_heat_error}.

\subsection{Heat-type equations in finance}
We consider a model of price formation initially proposed in Ref. \cite{lasry_mean_2007};
see also Refs. \cite{markowich_parabolic_2009,caffarelli_price_2011,burger_boltzmann-type_2013,achdou_partial_2014,pietschmann_partial_2012}. This model describes
the density of buyers $f_{{\text B}}(x,t)$ and the density of vendors
$f_{{\text V}}(x,t)$ in a system, as functions of of the bid or, respectively,
ask price $x\in\mathbb{R}$ for a certain good being traded between
them, and time $t\in[0,\infty)$.

The idea is that when a buyer and vendor agree on a price, the transaction
takes place; the buyer then becomes a vendor, and vice-versa. However,
it is also assumed that there exists a fixed transaction fee $a\in\mathbb{R}$.
Consequently, the actual buying price is $x+a$, and so the (former)
buyer will try to sell the good at the next trading event not for
the price $x$, but for $x+a$. Similarly, the profit for the vendor
is actually $x-a$, and so he/she would not be willing to pay more
than $x-a$ for the good at the next trading event. In time, this
system should achieve an equilibrium.

Mathematically, the dynamics of the buyer/vendor densities is assumed
to be governed by the heat equation with a certain source term. The
source term in each case is simply the (time-dependent) transaction
rate $\lambda(t)$, corresponding to the flux of buyers and vendors,
at the particular price where the trading event occurs, shifted accordingly
by the transaction cost. Thus the system is described by
\begin{equation}
\begin{cases}
\left(\partial_{t}-\partial_{x}^{2}\right)f_{{\text B}}=\lambda\left(t\right)\delta\left(x-\left(x_{p}\left(t\right)-a\right)\right)\,, & {\rm for}\,\,x<x_{p}\left(t\right)\,,\\
f_{{\text B}}=0\,, & {\rm for}\,\,x>x_{p}\left(t\right)\,,
\end{cases}
\end{equation}
and
\begin{equation}
\begin{cases}
\left(\partial_{t}-\partial_{x}^{2}\right)f_{{\text V}}=\lambda\left(t\right)\delta\left(x-\left(x_{p}\left(t\right)+a\right)\right)\,, & {\rm for}\,\,x>x_{p}\left(t\right)\,,\\
f_{{\text V}}=0\,, & {\rm for}\,\,x<x_{p}\left(t\right)\,,
\end{cases}
\end{equation}
where the free boundary $x_{p}(t)$ represents the agreed price of
trading at time $t$, and the transaction rate is $\lambda(t)=-\partial_{x}f_{{\text B}}(x_{p}(t),t)=\partial_{x}f_{{\text V}}(x_{p}(t),t)$.
(NB: The functional form of $\lambda(t)$ is uniquely fixed simply by the
requirement that the two densities are conserved, i.e. $\partial_{t}\int{\rm d}xf_{{\text B}}=0=\partial_{t}\int{\rm d}xf_{{\text V}}$,
under the assumption that we have homogeneous Neumann BCs at the left
and right boundaries respectively.) Now, we can actually combine this
system into a single problem for the difference between buyer and
vendor densities, 
\begin{equation}
f=f_{{\text B}}\Theta\left(-\left(x-x_{p}\left(t\right)\right)\right)-f_{{\text V}}\Theta\left(x-x_{p}\left(t\right)\right)\,.
\end{equation}
The ``spatial'' (i.e. price) domain can be taken to be bounded,
and homogeneous Neumann BCs are assumed at the boundaries. Thus the problem we are interested in is: 
\begin{equation}
\begin{cases}
\partial_{t}f-\partial_{x}^{2}f=\lambda\left(t\right)\left(\delta\left(x-x_{p_{-}}\left(t\right)\right)-\delta\left(x-x_{p_{+}}\left(t\right)\right)\right)\,, & x\in\mathscr{I}=\left[0,1\right],\enskip t>0\,,\\
f\left(x,0\right)=f_{\text{I}}\left(x\right),\enskip f\gtrless0\enskip{\rm for}\enskip x\lessgtr x_{p}\left(t\right)\,, & \partial_{x}f\left(0,t\right)=\partial_{x}f\left(1,t\right)=0\,.
\end{cases}\label{eq:pde_finance_model}
\end{equation}
where $\lambda(t)=-\partial_{x}f(x_{p}(t),t),$ and we have defined
$x_{p_{\pm}}(t)=x_{p}(t)\pm a$. Moreover, one can show that from
this setup, it follows that the free boundary evolves via
\begin{equation}
\dot{x}_{p}\left(t\right)=\frac{\partial_{x}^{2}f\left(x_{p}\left(t\right),t\right)}{\lambda\left(t\right)}\,.\label{eq:pde_finance_price_evolution}
\end{equation}

In this case, we have not one but two singular source locations on
the RHS of the PDE. Hence, in order to implement the PwP method, we
must here divide the spatial domain $\mathscr{I}$ into three disjoint
regions, with the two singularity locations at their interfaces: $\mathscr{I}=\mathscr{D}^{-}\cup\mathscr{D}^{0}\cup\mathscr{D}^{+}$
with $\mathscr{D}^{-}=[0,x_{p_{-}}(t)]$, $\mathscr{D}^{0}=[x_{p_{-}}(t),x_{p_{+}}(t)]$
and $\mathscr{D}^{+}=[x_{p_{+}}(t),1]$. Then, we decompose $f=f^{-}\Theta^{-}+f^{0}\Theta^{0}+f^{+}\Theta^{+}$
with $\Theta^{-}=\Theta(-(x-x_{p_{-}}(t)))$, $\Theta^{0}=\Theta(x-x_{p_{-}}(t))-\Theta(x-x_{p_{+}}(t))$
and $\Theta^{+}=\Theta(x-x_{p_{+}}(t))$. Inserting this into the
PDE (\ref{eq:pde_finance_model}), we get homogeneous problems $(\partial_{t}-\partial_{x}^{2})f^{\sigma}=0$
on $\mathscr{D}^{\sigma}$ for $\sigma\in\{0,\pm\}$, along with the
JCs $[f]_{p_{\pm}}=0$ and $[\partial_{x}f]_{p_{\pm}}=\pm\lambda(t)$.

Before proceeding to the numerical implementation, we note that it
is possible to derive an exact stationary (i.e. $t\rightarrow\infty$)
solution of the problem (\ref{eq:pde_finance_model}). In particular,
denoting the (time-conserved) number of buyers and vendors, respectively,
by $N_{{\text B}}=\int_{0}^{x_{p}}{\rm d}x\:f$ and $N_{{\text V}}=-\int_{x_{p}}^{1}{\rm d}x\:f$,
one can show that in the stationary ($t\rightarrow\infty$) limit,
\begin{align}
 & \begin{cases}
N_{{\text B}}=\, & -\lambda^{\text{stat}}a(x_{p}^{\text{stat}}-a/2)\,,\\
N_{{\text V}}=\, & -\lambda^{\text{stat}}a(1-x_{p}^{\text{stat}}-a/2)\,,
\end{cases}\\
\Leftrightarrow\, & \begin{cases}
\lambda^{\text{stat}}=\, & \left[-\left(N_{{\text B}}+N_{{\text V}}\right)\right]/\left[a\left(1-a\right)\right]\,,\\
x_{p}^{\text{stat}}=\, & \left[2N_{{\text B}}+a\left(N_{{\text V}}-N_{{\text B}}\right)\right]/\left[2\left(N_{{\text B}}+N_{{\text V}}\right)\right]\,,
\end{cases}\label{eq:pde_finance_stationary_lambda_price}
\end{align}
which we can use to determine the exact stationary solution
\begin{equation}
\lim_{t\rightarrow\infty}f\left(x,t\right)=f^{\text{stat}}\left(x\right)=\begin{cases}
-\lambda^{\text{stat}}a\,, & {\rm for}\,\,0\leq x<x_{p_{-}}^{\text{stat}}\,,\\
\lambda^{\text{stat}}\left(x-x_{p}^{\text{stat}}\right)\,, & {\rm for}\,\,x_{p_{-}}^{\text{stat}}\leq x\leq x_{p_{+}}^{\text{stat}}\,,\\
\lambda^{\text{stat}}a\,, & {\rm for}\,\,x_{p_{+}}^{\text{stat}}<x\leq1\,.\label{eq:pde_finance_stationary_solution}
\end{cases}
\end{equation}

The problem (\ref{eq:pde_finance_model}) is solved numerically in Ref. \cite{markowich_parabolic_2009}
(see also section 2.5.2 of \cite{pietschmann_partial_2012}) using (Gaussian) delta function approximations for the source on an equispaced computational grid. We implement here using our PwP method the exact same setup: in particular, we take
a transaction fee of $a=0.1$ and initial data $f_{\text{I}}(x)=\tfrac{875}{6}x^{3}-\tfrac{700}{3}x^{2}+\tfrac{175}{2}x$.
(NB: Despite the fact that this does not actually satisfy homogeneous
Neumann BCs, the numerical evolution will force it to.) Analytically,
we have $x_{p}(0)=\tfrac{3}{5}$ and $\lambda(0)=35$. Also, using (\ref{eq:pde_finance_stationary_lambda_price}), we have $\lambda^{\text{stat}}=-\frac{8855}{162}$
and $x_{p}^{\text{stat}}=\frac{731}{1012}\approx0.7223$. As we evolve
forward in time, we use Chebyshev polynomial interpolation to determine
the transaction rate $\lambda(t)$ (i.e. the negative of the spatial
derivative of the solution at $x_{p}(t)$) as well as the evolution
of $x_{p}(t)$ via (\ref{eq:pde_finance_price_evolution}).

\begin{figure}
\begin{center}
\includegraphics[scale=0.6]{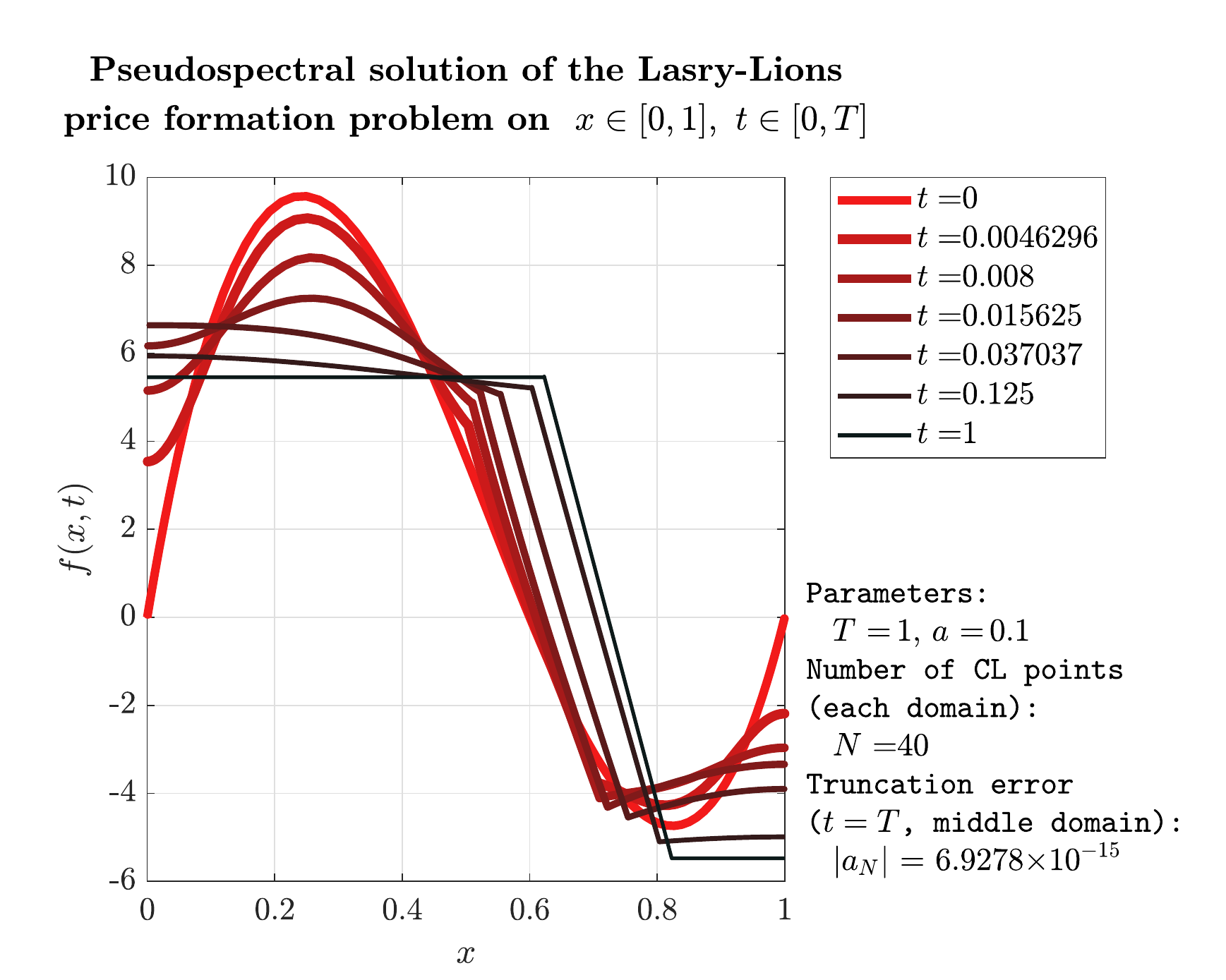}
\includegraphics[scale=0.6]{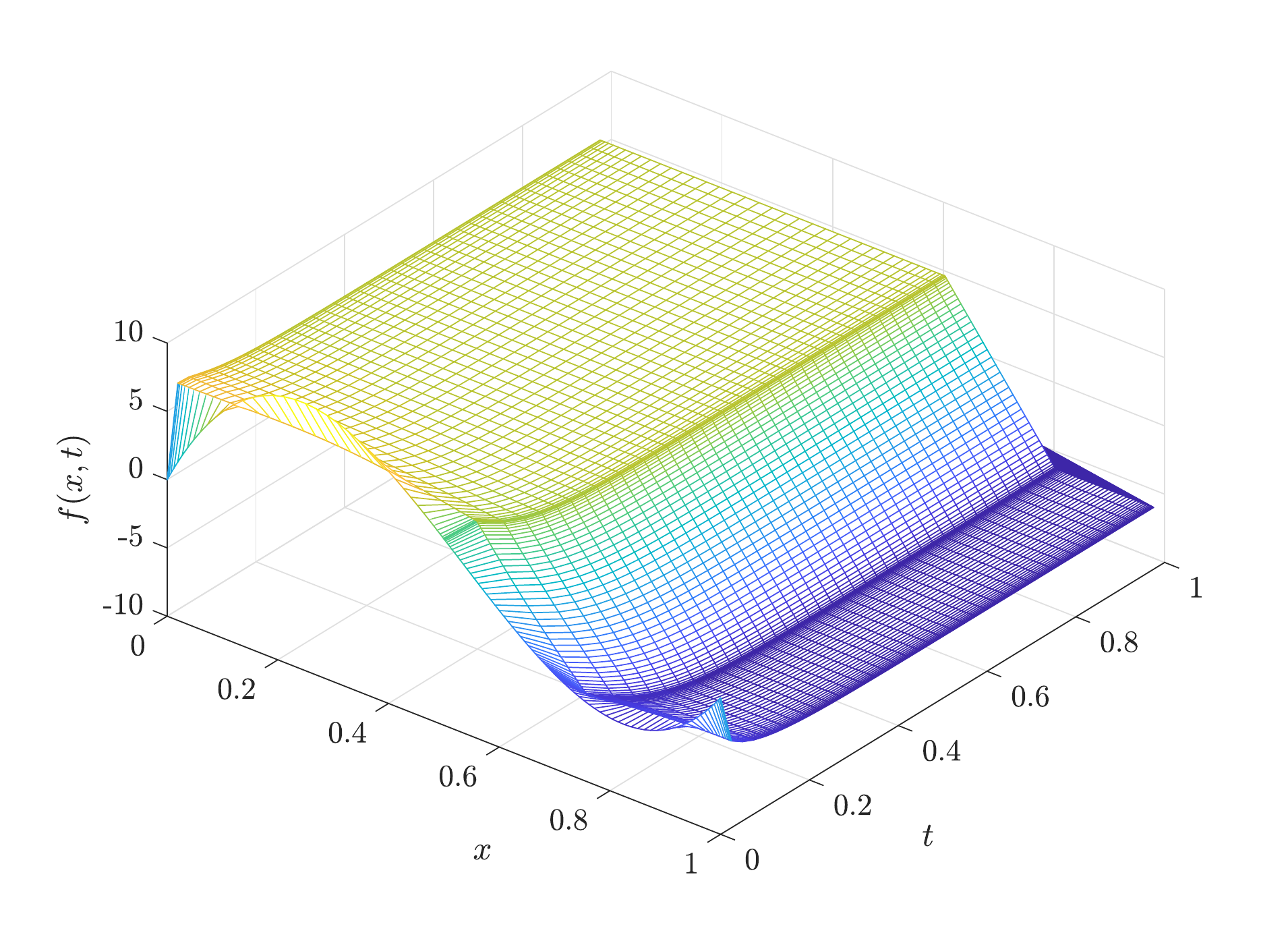}
\caption{Solution of the problem (\ref{eq:pde_finance_model}).}\label{fig:pde_pwp_price_formation_soln}
\end{center}
\end{figure}

\begin{figure}
\begin{center}
\includegraphics[scale=0.6]{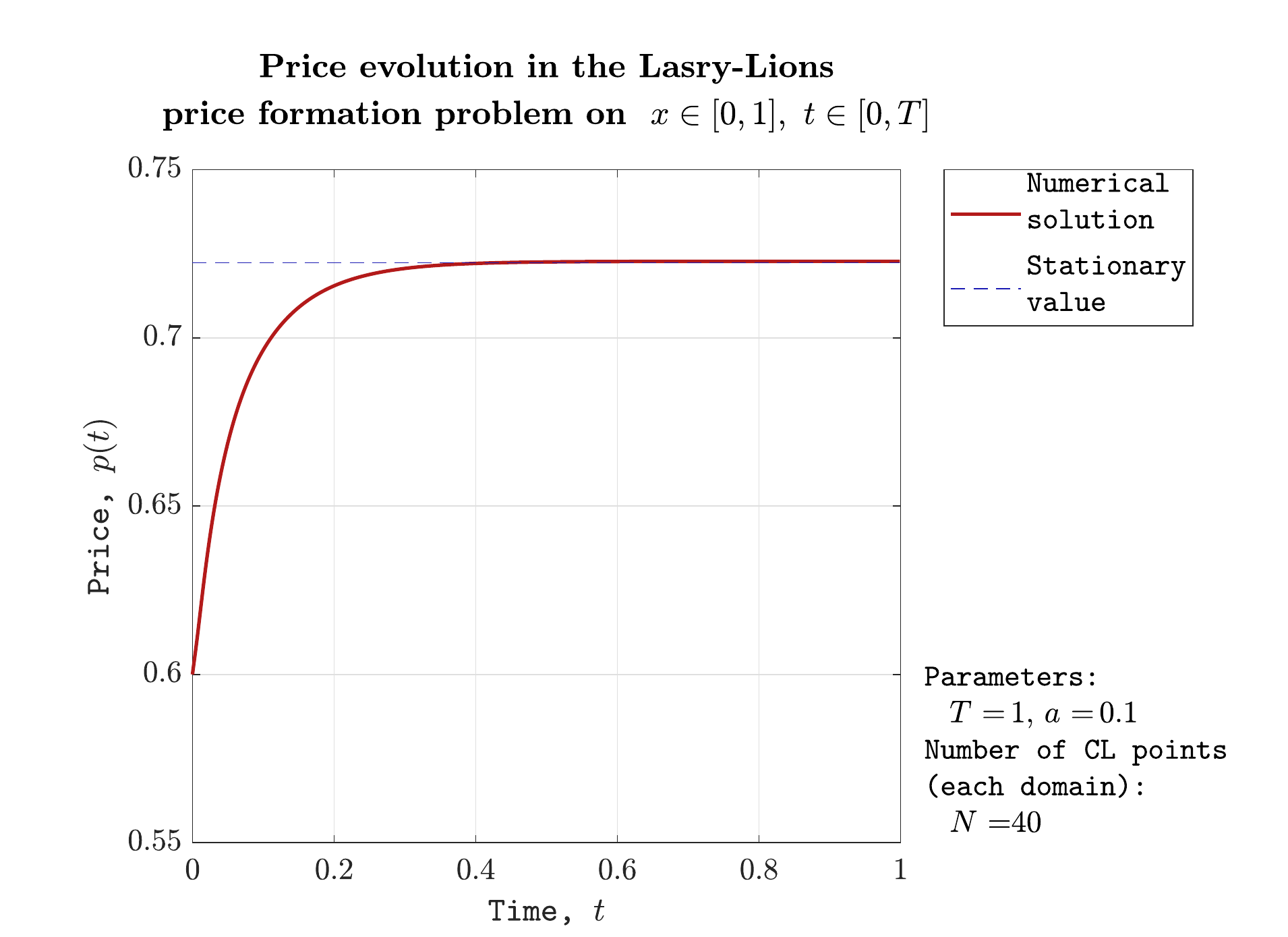}
\includegraphics[scale=0.6]{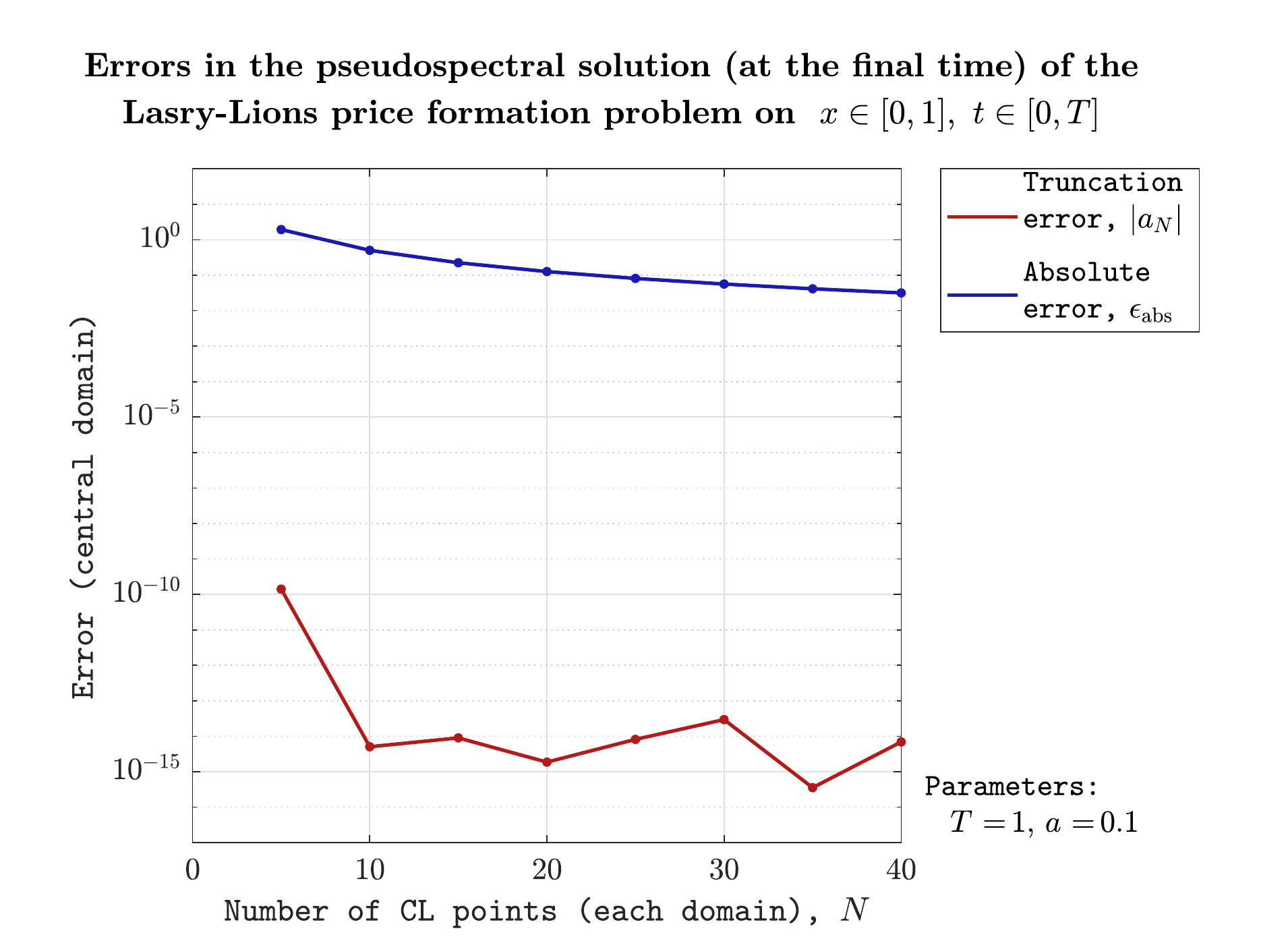}
\caption{Price evolution and convergence of the numerical scheme for the problem (\ref{eq:pde_finance_model}).}\label{fig:pde_pwp_price_formation_price_and_error}
\end{center}
\end{figure}

The numerical scheme is given in Appendix \ref{a-numerical-parabolic},
and results in Figures \ref{fig:pde_pwp_price_formation_soln} and \ref{fig:pde_pwp_price_formation_price_and_error}. In particular, in Figure \ref{fig:pde_pwp_price_formation_soln} we show the numerical solution for $f$, and in Figure \ref{fig:pde_pwp_price_formation_price_and_error}, the price as a function of time as well as the numerical convergence rates. For the latter, we plot not only the truncation error but also the absolute error with the stationary solution (\ref{eq:pde_finance_stationary_solution}), in this case, using the infinity norm: $\epsilon_{\text{abs}}=||\bm{f}-\bm{f}^{\text{stat}}||_{\infty}$. Of course, since we can only evolve the solution up to a finite time (which we choose to be $t=T=1$), we should not expect this to converge to zero; however, its decline with increasing $N$ nevertheless serves to illustrate a good validation of our results.

We remark that our numerical implementation here not only requires an order of magnitude fewer grid points than that of Ref. \cite{markowich_parabolic_2009}, but in fact yields convergence to the \emph{correct} stationary solution while that of Ref. \cite{markowich_parabolic_2009} \emph{does not}. Indeed, in the latter, not only are more points required (essentially due to the necessity of resolving well enough the Gaussian-approximated delta functions) but the scheme actually fails, even so, to approach (\ref{eq:pde_finance_stationary_solution}) as well as ours by the same finite time, $t=T=1$. (To wit, Ref. \cite{markowich_parabolic_2009} obtains $x_{p}\rightarrow0.71$ in the large $t$ limit, instead of the correct value, $0.7223$, which we achieve with our PwP method as shown in Figure \ref{fig:pde_pwp_price_formation_price_and_error}.)

\section{Second order hyperbolic PDEs}
\label{second-order-hyperbolic-pdes}

We move on to consider in this section second order hyperbolic problems.
In particular, we first solve the standard $(1+1)$-dimensional elastic
wave equation, taking a delta derivative source. Afterwards, we discuss
possible physical applications of this and obstacles thereto---including
problems in gravitational physics and seismology. 

\subsection{Wave equation}

Let us consider the the elastic wave equation, in the form of the
following simplified $(1+1)$-dimensional problem for $u(x,t)$ with
a delta function derivative source at a fixed point $x_{*}\in\mathscr{I}=[0,L]$,
and homogeneous Dirichlet boundary conditions:
\begin{equation}
\begin{cases}
\partial_{t}^{2}u-\partial_{x}^{2}u=g\left(t\right)\delta'\left(x-x_{*}\right)\,, & x\in\mathscr{I}=\left[0,L\right]\,,\enskip t>0\,,\\
u\left(x,0\right)=0\,,\enskip\partial_{t}u\left(x,0\right)=0\,, & u\left(0,t\right)=0=u\left(L,t\right)\,.
\end{cases}\label{eq:elastic_wave_pde}
\end{equation}

It is actually possible to derive an exact solution for this problem
on an unbounded domain $\mathscr{I}=\mathbb{R}$. For the interested
reader, the procedure is explained in Appendix \ref{a-solution-elastic-wave}.
For concreteness we take a simple sinusoidal source time function
$g(t)=\kappa\sin(\omega t)$, in which case the exact solution reads:
\begin{align}
u_{{\text{ex}}}\left(x,t\right)=\, & \kappa\bigg[\frac{1}{4}\sum_{\sigma=\pm}\sigma{\rm sgn}\left(x-x_{*}+\sigma t\right)\sin\left(\omega\left(x-x_{*}+\sigma t\right)\right)\nonumber \\
 & -\frac{1}{2}{\rm sgn}\left(x-x_{*}\right)\cos\left(\omega\left(x-x_{*}\right)\right)\sin\left(\omega t\right)\bigg]\,,\label{eq:elastic_wave_solution_exact}
\end{align}
where ${\rm sgn}(\cdot)$ is the sign function, with the property
${\rm d}({\rm sgn}(x))/{\rm d}x=2{\rm d}\Theta(x)/{\rm d}x=2\delta(x)$.

To solve (\ref{eq:elastic_wave_pde}) numerically, we implement the
now familiar PwP decomposition: $u=u^{-}\Theta^{-}+u^{+}\Theta^{+}$
where $\Theta^{\pm}=\Theta(\pm(x-x_{*}))$. Inserting this into (\ref{eq:elastic_wave_pde}),
we get homogeneous PDEs $\partial_{t}^{2}u^{\pm}-\partial_{x}^{2}u^{\pm}=0$
to the left and right of the singularity, $x\in\mathscr{D}^{-}=[0,x_{*}]$
and $x\in\mathscr{D}^{+}=[x_{*},L]$ respectively, along with the
JCs $[u]_{p}=-g(t)$ and $[\partial_{x}u]_{p}=0$. We now proceed
by recasting (\ref{eq:elastic_wave_pde}) as a first-order hyperbolic
system for $\vec{U}=[u\enskip v\enskip w]^{{\rm T}}$ with $v=\partial_{x}u$
and $w=\partial_{t}u$, as
\begin{equation}
\partial_{t}\vec{U}=\left[\begin{array}{ccc}
0 & 0 & 0\\
0 & 0 & 1\\
0 & 1 & 0
\end{array}\right]\partial_{x}\vec{U}+\left[\begin{array}{ccc}
0 & 0 & 1\\
0 & 0 & 0\\
0 & 0 & 0
\end{array}\right]\vec{U}\quad{\rm on}\enskip\mathscr{D}^{\pm}\,,\quad\left[\vec{U}\right]_{p}=\left[\begin{array}{c}
-g\\
0\\
-\dot{g}
\end{array}\right]\,.\label{eq:elastic_wave_pde_first-order}
\end{equation}

\begin{figure}
\begin{center}
\includegraphics[scale=0.6]{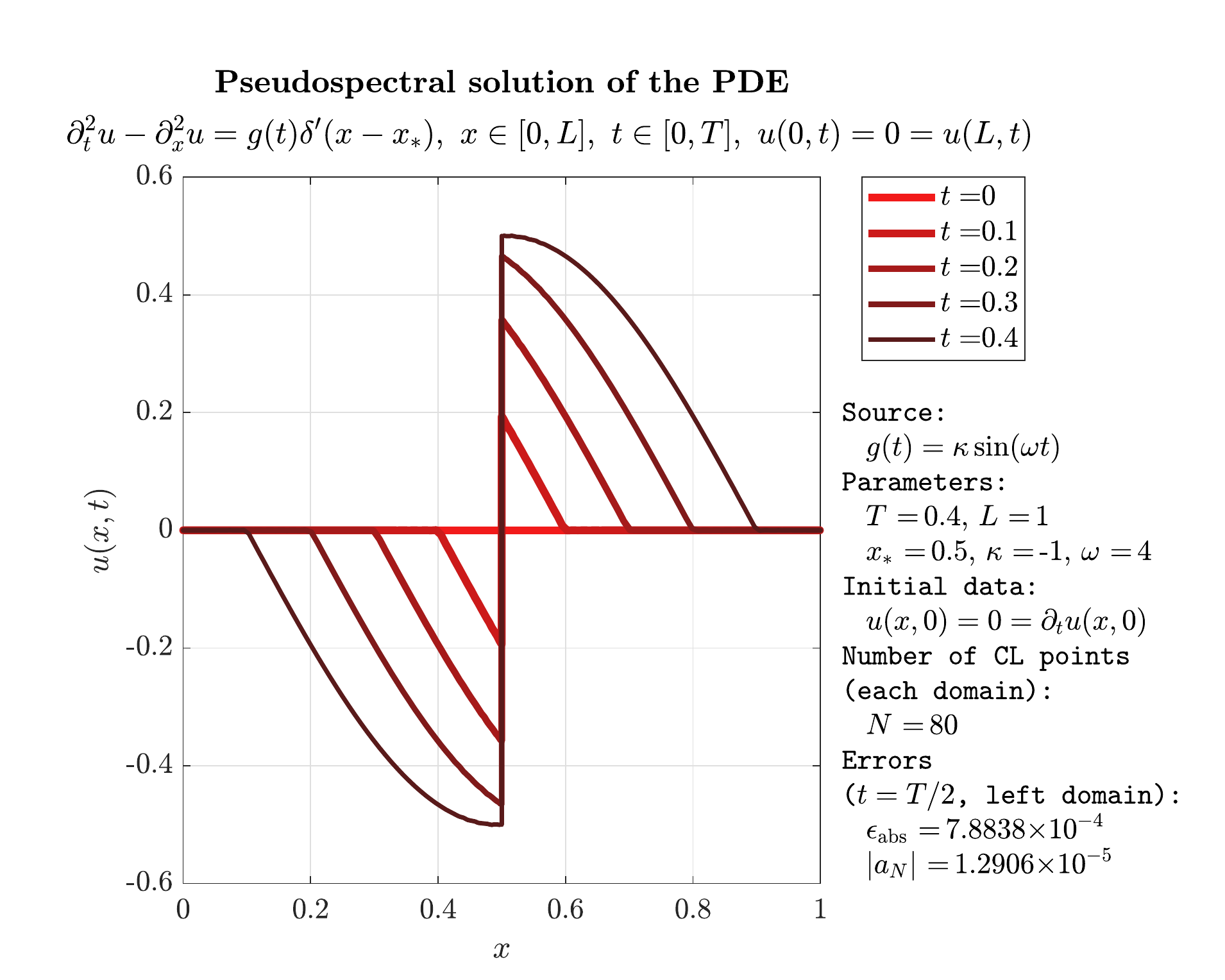}
\includegraphics[scale=0.6]{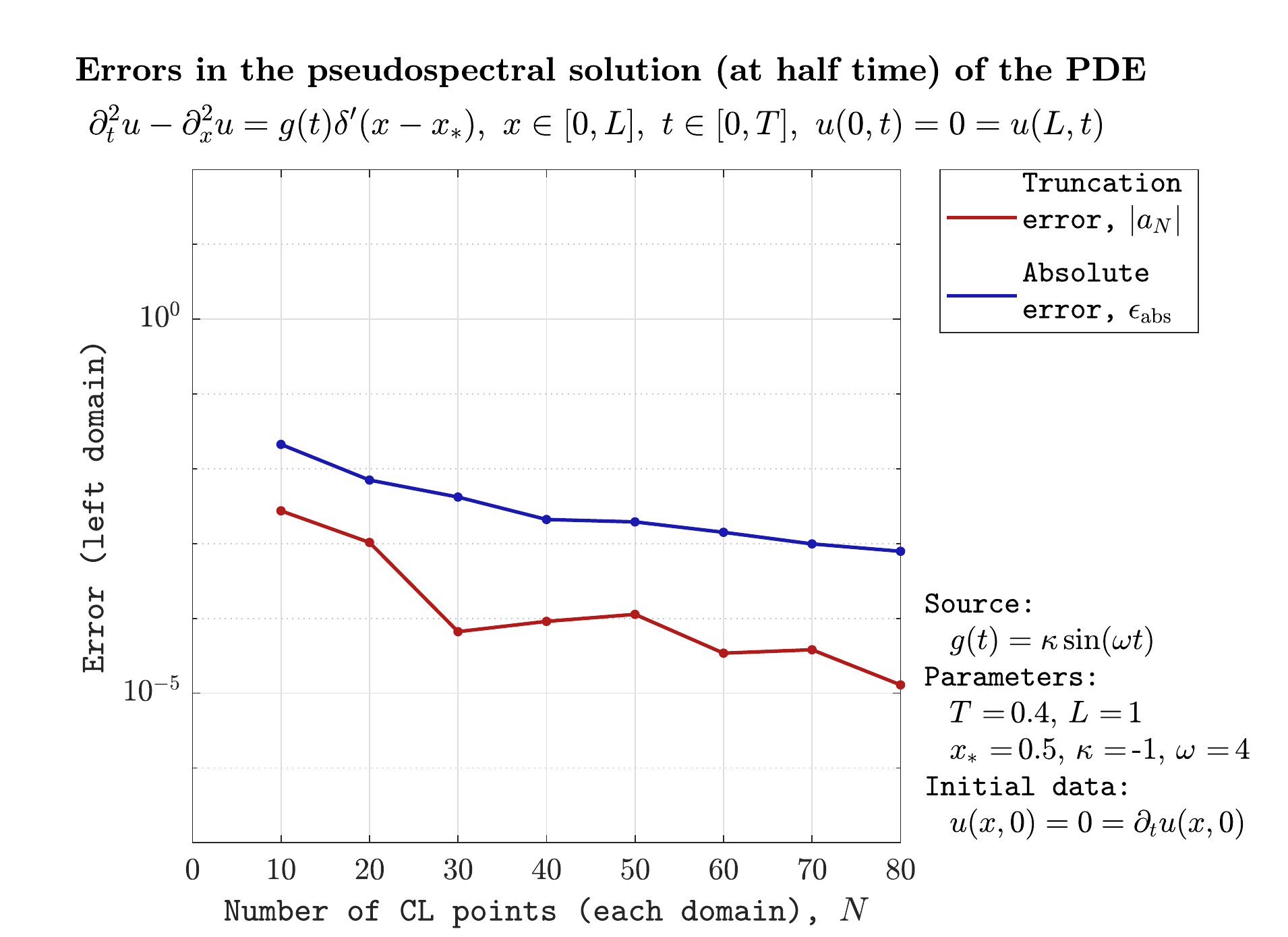}
\caption{Solution and convergence of the numerical scheme for the problem (\ref{eq:elastic_wave_pde}).}\label{fig:pde_pwp_wave_soln_and_error}
\end{center}
\end{figure}

\begin{figure}
\begin{center}
\includegraphics[scale=0.6]{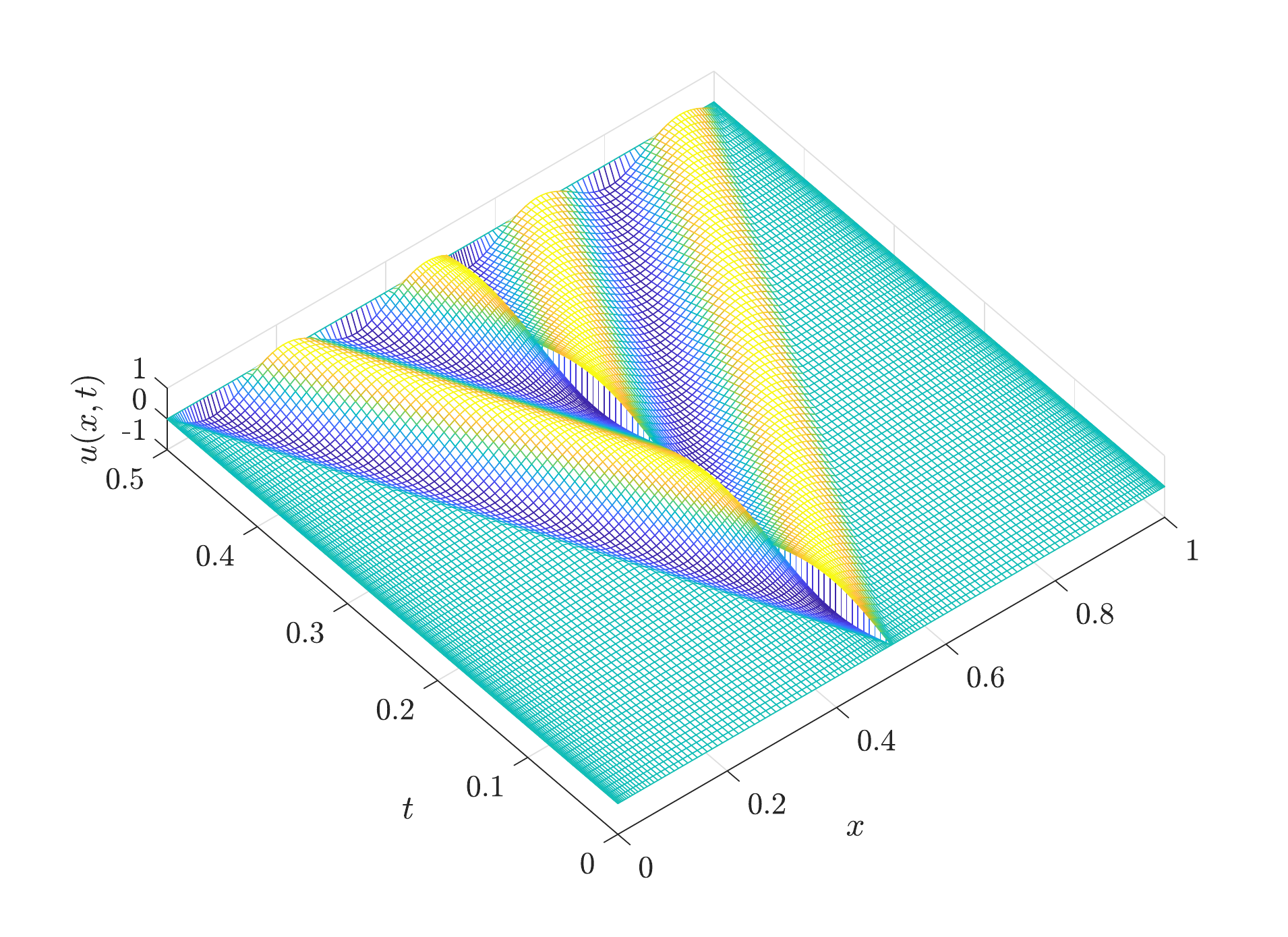}
\caption{Solution (with $N=80$) of the same problem as in Figure \ref{fig:pde_pwp_wave_soln_and_error} but using $\omega=24$.}\label{fig:pde_pwp_wave_soln3d}
\end{center}
\end{figure}

The numerical scheme is given in Appendix \ref{a-numerical-second-order-hyperbolic},
and results in Figures \ref{fig:pde_pwp_wave_soln_and_error} and \ref{fig:pde_pwp_wave_soln3d}. The absolute error is again computed in the infinity norm on the CL grids: $\epsilon_{\text{abs}}=||\bm{u}-\bm{u}_{\text{ex}}||_{\infty}$.

The same problem (\ref{eq:elastic_wave_pde}) is considered numerically in Ref. \cite{petersson_stable_2010}, but using a different (polynomial) source function $g(t)$, and a discretization procedure for the delta function (derivatives) on the computational grid (carried out in such a way that the distributional action thereof yields the expected result on polynomials up to a given degree). With our PwP method here, we obtain the same order of magnitude of the (absolute) error in the numerical solution as that in Ref. \cite{petersson_stable_2010} for the same (order of magnitude of) number of grid points; however the drawback of the ``discretized delta'' method of Ref. \cite{petersson_stable_2010}, in contrast to the PwP method, is that the solution in the former is visibly quite poorly resolved close to the singularity.

We add that we have also carried out the solution to the problem shown in Figure \ref{fig:pde_pwp_wave_soln_and_error} using higher-order (from second up to eighth order) finite-difference time evolution schemes. These yield no visible improvement (at any order tried) in either the absolute or the truncation error relative to the first-order time evolution results. Thus the spacial pseudospectral grid appears to control the total level of the error, with a higher-order scheme for the time evolution producing, at least in this case, no greater benefits.

\subsection{Wave-type equations in physical applications}

As we have mentioned, our numerical
studies in this paper are largely motivated by their applicability
to the computation of the self-force in gravitational physics. There
one encounters different levels of complexity of this problem, the
simplest being that of the self-force due to a scalar field---as
a conceptual testbed for the more complicated and realistic problem
of the full self-force of the gravitational field---in a fixed (non-dynamical) black
hole spacetime. This can refer to a non-spinning (Schwarzschild-Droste\footnote{Most commonly, this is referred to simply as the ``Schwarzschild solution'' in general relativity. Yet, it has long gone largely unrecognized that Johannes Droste, then a doctoral student of Lorentz, discovered this solution independently and announced it only four months after Schwarzschild \cite{dro16a,dro16b,sc16,ro02}, so for the sake of historical fairness, we here use the nomenclature ``Schwarzschild-Droste solution'' instead.}) black
hole, where the problem is of the form (\ref{eq:intro_general_pde})
where $\mathcal{L}=\partial_{t}^{2}-\partial_{x}^{2}+V$ is just a
simple $(1+1)$-dimensional wave operator with some known potential
$V$ and a source $S=f\delta_{(p)}$, with $\dim(\mathscr{I})=1$.
We recognize this now as quite typical for the application of our PwP
method, and indeed this has been done with success in the past \cite{canizares_extreme-mass-ratio_2011,canizares_efficient_2009,canizares_pseudospectral_2010,canizares_time-domain_2011,canizares_tuning_2011,jaramillo_are_2011,canizares_overcoming_2014,oltean_frequency-domain_2017}. As we briefly remarked in the Introduction, the main difference between most of these works and our numerical schemes throughout this paper is that for the time evolution, rather than relying on finite-difference methods, the former made use of the method of lines. This can be quite well-suited especially for these types of $(1+1)$-dimensional hyperbolic problems, which can be formulated in terms of characteristic fields propagating along the two lightcone directions ($t\pm x = {\textrm{const.}}$). The imposition of the JCs is then achieved quite simply in this setting by just evolving, in the left domain, the characteristic field propagating towards the right and relating it (via the JC) to the value of the characteristic field propagating towards the left in the right domain. For the interested reader, this kind of procedure is described in detail in Chapter 3 of Ref. \cite{canizares_extreme-mass-ratio_2011}.

We could also consider the scalar self-force problem in a spinning (Kerr)
black hole spacetime, however the issue there---owing to the existence of fewer symmetries in the problem than in the non-spinning case---is that $\dim(\mathscr{I})=2$
in the time domain (with a more complicated second-order hyperbolic
operator $\mathcal{L}$); however, this could be remedied for a possible PwP implementation by passing to the
frequency domain, which transforms (\ref{eq:intro_general_pde}) to
an ODE (with $\dim(\mathscr{I})=1$, $\mathscr{V}=\emptyset$, and
again, a simple source $S=f\delta_{(p)}$).

The application of the PwP method to the full gravitational self-force
is a subject of ongoing work, however (modulo certain technical problems
relating to the gauge choice, which we will not elaborate upon here)
in the Schwarzschild-Droste case it essentially reduces to solving the same
type of problem (\ref{eq:intro_general_pde}) with $\dim(\mathscr{I})=1$
and $S=f\delta_{(p)}+g\delta'_{(p)}$. The equivalent problem in the
Kerr case once again suffers from the issue that $\dim(\mathscr{I})=2$
in the time domain, so the PwP method cannot be applied there except
after a transformation to the frequency domain (which produces $\dim(\mathscr{I})=1$
and $S=f\delta_{(p)}+g\delta'_{(p)}+h\delta''_{(p)}$ in this case).

Outside of gravitational physics, another setting where the PwP technique
could also possibly prove useful is in seismology. There, however,
the modeling of seismic waves \cite{romanowicz_seismology_2007,aki_quantitative_2009,shearer_introduction_2009,madariaga_seismic_2007,petersson_stable_2010} typically involves equations of the
form (\ref{eq:pwp-multivariable_source_Lu}) with $3$-dimensional
delta functions (i.e. $\dim(\mathscr{I})=\bar{n}=3$, usually referring
to the $3$ dimensions of ordinary space) which, as we have amply
discussed in relation thereto, are not directly amenable to a PwP-type
approach as such. However, the methods outlined in this paper might
be of some use if symmetries or other simplifying assumptions can,
in a situation of interest, reduce the dimension of the distributional
source to $1$ (as an alternative to delta function approximation
procedures, which are common practice in this area as well).

\section{Elliptic PDEs}
    \label{elliptic-pdes}

Finally, we consider in this section the elliptical problem appearing
in section 4.3 of Ref. \cite{tornberg_numerical_2004}: namely, the Poisson equation on a square
of side length $2$ centered on the origin in $\mathbb{R}^{2}$, with
a simple (negative) one-dimensional delta function source supported
on the circle of radius $r_{*}=\tfrac{1}{2}$,
\begin{equation}
\begin{cases}
\triangle_{\mathbb{R}^{2}}u=-\delta\left(r-r_{*}\right)\,, & \text{on }\mathscr{U}=\left[-1,1\right]\times\left[-1,1\right]\subset\mathbb{R}^{2}\,,\\
u=1-\tfrac{1}{2}\log\left(2r\right)\,, & \text{on }\partial\mathscr{U}\,.
\end{cases}\label{eq:poisson_pde}
\end{equation}
In this case, the polar symmetry of the PDE entails that the solution
will only depend on the radial coordinate $r$ (which in this case
notationally substitutes the $x$ coordinate in antecedent sections).
Indeed, (\ref{eq:poisson_pde}) has an exact solution which is simply
given by
\begin{equation}
u_{\text{ex}}=1-\tfrac{1}{2}\log\left(2r\right)\Theta\left(r-r_{*}\right)\,.\label{eq:poisson_solution_exact}
\end{equation}

We can use the fact that in polar coordinates, $\triangle_{\mathbb{R}^{2}}=\partial_{r}^{2}+\tfrac{1}{r}\partial_{r}+\tfrac{1}{r^{2}}\partial_{\theta}^{2}$,
and so numerically all we need to do is solve $(\partial_{r}^{2}+\tfrac{1}{r}\partial_{r})u(r)=-\delta(r-r_{*})$
for a given $\theta\in[0,2\pi]$, where the value of $\theta$ will
determine $\{r\}=\mathscr{I}=[0,L]$ and hence the BC at $u(L)$ (that
is, on $\partial\mathscr{U}$), and repeat over some set of discrete
$\theta$ values in case the entire numerical solution on the $(r,\theta)$-plane
is desired.

Thus, we simply implement the PwP method here by writing $u=u^{-}\Theta^{-}+u^{+}\Theta^{+}$
for $\Theta^{\pm}=\Theta(r-r_{*})$, whereby we obtain the homogeneous
equations $(\partial_{r}^{2}+\tfrac{1}{r}\partial_{r})u^{\pm}=0$
along with the JCs $[u]_{*}=0$ and $[\partial_{r}u]_{*}=-1$.

\begin{figure}
\begin{center}
\includegraphics[scale=0.6]{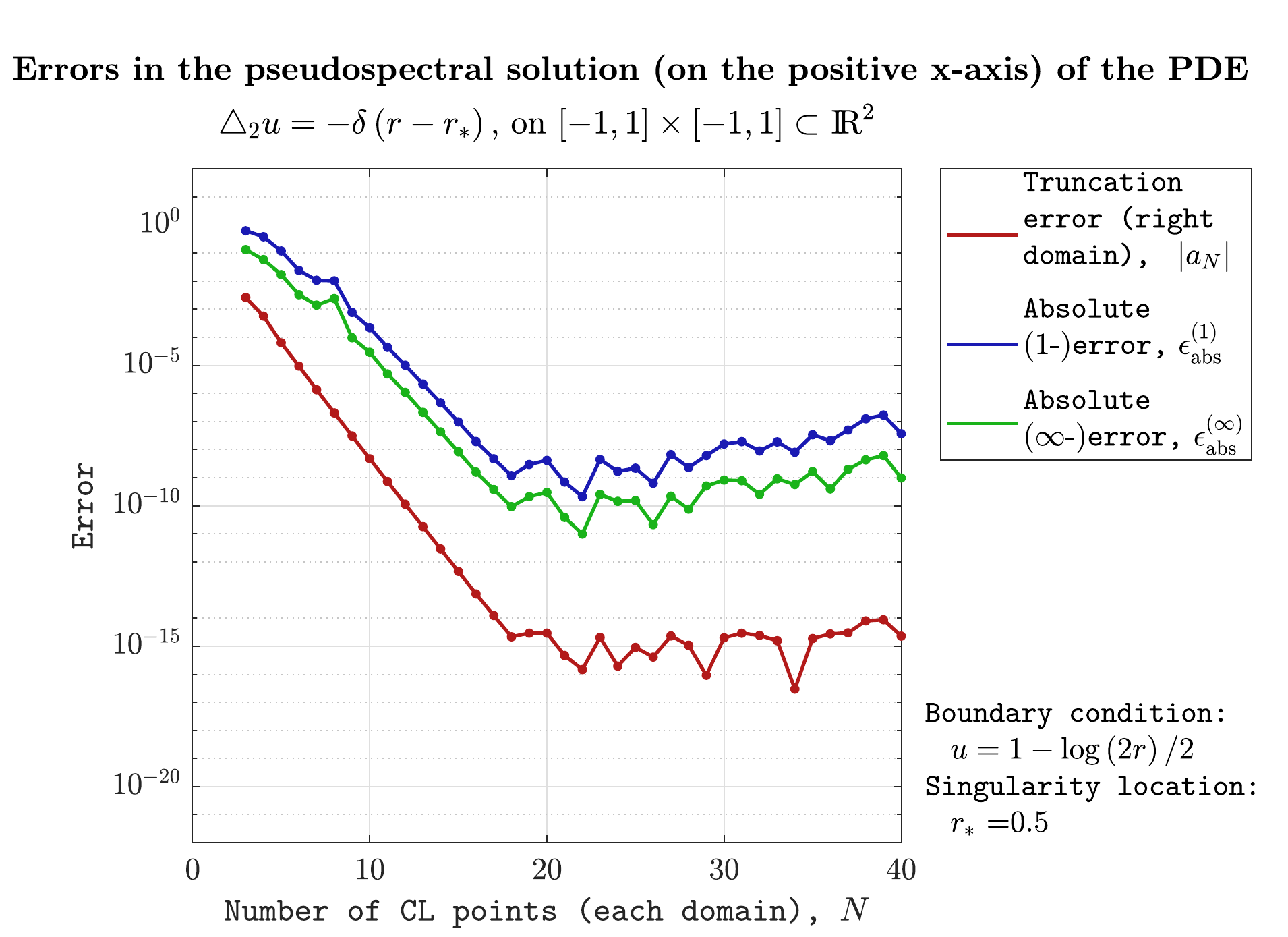}
\caption{Convergence of the pseudospectral numerical scheme for the problem (\ref{eq:poisson_pde}).}\label{fig:pde_pwp_poisson_error}
\end{center}
\end{figure}

The detailed numerical scheme is given in Appendix \ref{a-numerical-elliptic}, and results in
Figure \ref{fig:pde_pwp_poisson_error}. In this case, we simply plot the errors along the positive
$x$-axis in $\mathbb{R}^{2}$ on the CL grids: in addition to the
right-domain truncation error, we also show (as is done in Ref. \cite{tornberg_numerical_2004})
the absolute error in both the $l^{1}$-norm, $\epsilon_{\text{abs}}^{(1)}=||\bm{u}-\bm{u}_{\text{ex}}||_{1}$,
as well as in the infinity norm, $\epsilon_{\text{abs}}^{(\infty)}=||\bm{u}-\bm{u}_{\text{ex}}||_{\infty}$.
Up to $N\approx20$, we observe the typical (exponential) spectral
convergence of all three errors, with a significant (by a few orders
of magnitude) improvement over the results of Ref. \cite{tornberg_numerical_2004} (using delta function
approximations) for the latter two.

\section{Conclusions}
    \label{conclusions}

We have expounded in this paper a practical approach---the ``Particle-without-Particle'' (PwP) method---for numerically solving differential equations with distributional sources; to summarize, one does this by breaking up the solution into (regular function) pieces supported between---plus, if necessary, at---singularity (``particle'') locations, solving sourceless (homogeneous) problems for these pieces, and then matching them via the appropriate ``jump'' (boundary) conditions effectively substituting the original singular source. Building upon its successful prior application in the specific context of the self-force problem in general relativity, we have here generalized this method and have shown it to be viable for any linear partial differential equation of arbitrary order, with the provision that the distributional source is supported only on a one-dimensional subspace of the total problem domain. Accordingly, we have demonstrated its usefulness by solving first and second order hyperbolic problems, with applications in neuroscience and acoustics, respectively; parabolic problems, with applications in finance; and finally a simple elliptic problem. In particular, the numerical schemes we have employed for carrying these out have been based on pseudospectral collocation methods on Chebyshev-Lobatto grids. Generally speaking, our results have yielded varying degrees of improvement in the numerical convergence rates relative to other methods in the literature that have been attempted for solving these problems (typically relying on delta function approximation procedures on the computational grid).

We stress once more that the main limitations of the our PwP method as developed here are that it is only applicable to \emph{linear} problems with \emph{one}-dimensionally supported distributional sources. Thus, interesting lines of inquiry for future work might be to explore---however/if at all possible---extensions or adaptations of these ideas (a) to nonlinear PDEs, which would require working with nonlinear theories of distributions (having potential applicability to problems such as, e.g., the shallow-water equations with discontinuous bottom topography); (b) to more complicated sources than the sorts considered in this paper, perhaps even containing higher-dimensional distributions but possibly also requiring additional assumptions, such as symmetries (which might be useful for problems such as, e.g., seismology models with three-dimensional delta function sources).

\section*{Acknowledgements}

Due by MO to the Natural Sciences and Engineering Research Council
of Canada, by MO and ADAMS to LISA France-CNES, and by MO and CFS to the Ministry of Economy
and Competitivity of Spain, MINECO, contracts ESP2013-47637-P, ESP2015-67234-P and ESP2017-90084-P.

\appendix

\section{Proof of distributional identity}
\label{a-proof-identity}

The base case ($n=0$) reads $f(x,\mathbf{y})\delta(x-x_{p})\equiv f_{p}\delta(x-x_{p})$.
It is trivial to see that this holds.

Now assume the identity holds for $n=k$. Then we must prove that
it holds for $n=k+1$. In other words, we wish to show that: 
\begin{equation}
\left\langle f\delta_{\left(p\right)}^{\left(k+1\right)},\phi\right\rangle =\left\langle \left(-1\right)^{n}\sum_{j=0}^{n}\left(-1\right)^{j}\binom{n}{j}f_{p}^{\left(n-j\right)}\delta_{\left(p\right)}^{\left(j\right)},\phi\right\rangle .
\end{equation}
We begin with the LHS, and compute: 
\begin{align}
\left\langle f\delta_{\left(p\right)}^{\left(k+1\right)},\phi\right\rangle =\, & \int_{\mathscr{I}}{\rm d}x\,f\left(x,\mathbf{y}\right)\delta_{\left(p\right)}^{\left(k+1\right)}\left(x\right)\phi\left(x\right)\\
=\, & -\int_{\mathscr{I}}{\rm d}x\,\left(\phi'\left(x\right)f\left(x,\mathbf{y}\right)+\phi\left(x\right)f^{'}\left(x,\mathbf{y}\right)\right)\delta_{\left(p\right)}^{\left(k\right)}\left(x\right),
\end{align}
using integration by parts. Now, inserting the induction hypothesis,
\begin{align}
\left\langle f\delta_{\left(p\right)}^{\left(k+1\right)},\phi\right\rangle =\, & -\int_{\mathscr{I}}{\rm d}x\,\bigg\{\phi'\left(x\right)\left(-1\right)^{k}\sum_{j=0}^{k}\left(-1\right)^{j}\binom{k}{j}f_{p}^{\left(k-j\right)}\delta_{\left(p\right)}^{\left(j\right)}\left(x\right)\nonumber \\
 & +\phi\left(x\right)\left(-1\right)^{k}\sum_{l=0}^{k}\left(-1\right)^{l}\binom{k}{l}f_{p}^{\left(k+1-l\right)}\delta_{\left(p\right)}^{\left(j\right)}\left(x\right)\bigg\}\\
=\, & \left(-1\right)^{k+1}\bigg\{\sum_{j=0}^{k}\left(-1\right)^{j}\binom{k}{j}f_{p}^{\left(k-j\right)}\int_{\mathscr{I}}{\rm d}x\,\phi'\left(x\right)\delta_{\left(p\right)}^{\left(j\right)}\left(x\right)\nonumber \\
 & +\sum_{l=0}^{k}\left(-1\right)^{l}\binom{k}{l}f_{p}^{\left(k+1-l\right)}\int_{\mathscr{I}}{\rm d}x\,\phi\left(x\right)\delta_{\left(p\right)}^{\left(l\right)}\left(x\right)\bigg\}
\end{align}
Observe that, via integration by parts, $\langle\delta_{\left(p\right)}^{\left(j\right)},\phi\rangle=(-1)^{j}\phi_{p}^{\left(j\right)}$.
Hence, the above simplifies to: 
\begin{align}
\left\langle f\delta_{\left(p\right)}^{\left(k+1\right)},\phi\right\rangle =\, & \left(-1\right)^{k+1}\bigg\{\sum_{j=0}^{k}\binom{k}{j}f_{p}^{\left(k-j\right)}\phi_{p}^{\left(j+1\right)}+\sum_{l=0}^{k}\binom{k}{l}f_{p}^{\left(k+1-l\right)}\phi_{p}^{\left(l\right)}\bigg\}\\
=\, & \left(-1\right)^{k+1}\bigg\{\sum_{j=1}^{k+1}\binom{k}{j-1}f_{p}^{\left(k+1-j\right)}\phi_{p}^{\left(j\right)}+\sum_{l=0}^{k}\binom{k}{l}f_{p}^{\left(k+1-l\right)}\phi_{p}^{\left(l\right)}\bigg\}\\
=\, & \left(-1\right)^{k+1}\bigg\{ f_{p}^{\left(k+1\right)}\phi_{p}+\sum_{j=1}^{k+1}\left[\binom{k}{j-1}+\binom{k}{j}\right]f_{p}^{\left(k+1-j\right)}\phi_{p}^{\left(j\right)}+f_{p}\phi_{p}^{\left(k+1\right)}\bigg\}\label{eq:a-proof-sum_rearrangement}
\end{align}
after rearranging the sum terms. Using the recursive formula for the
binomial coefficient, $\binom{k}{j-1}+\binom{k}{j}=\binom{k+1}{j}$,
and then including the first and last terms in (\ref{eq:a-proof-sum_rearrangement})
into the sum, we get: 
\begin{equation}
\left\langle f\delta_{\left(p\right)}^{\left(k+1\right)},\phi\right\rangle =\left(-1\right)^{k+1}\sum_{j=0}^{k+1}\binom{k+1}{j}f_{p}^{\left(k+1-j\right)}\phi_{p}^{\left(j\right)}.
\end{equation}
Now using $\phi_{p}^{\left(j\right)}=\langle\delta_{\left(p\right)},\phi^{\left(j\right)}\rangle=(-1)^{j}\langle\delta_{\left(p\right)}^{\left(j\right)},\phi\rangle$,
we finally obtain 
\begin{align}
\left\langle f\delta_{\left(p\right)}^{\left(k+1\right)},\phi\right\rangle =\, & \int_{\mathscr{I}}{\rm d}x\,\phi\left(x\right)\bigg[\left(-1\right)^{k+1}\sum_{j=0}^{k+1}\left(-1\right)^{j}\binom{k+1}{j}f_{p}^{\left(k+1-j\right)}\delta_{\left(p\right)}^{\left(j\right)}\left(x\right)\bigg]\\
=\, & \left\langle \left(-1\right)^{k+1}\sum_{j=0}^{k+1}\left(-1\right)^{j}\binom{k+1}{j}f_{p}^{\left(k+1-j\right)}\delta_{\left(p\right)}^{\left(j\right)},\phi\right\rangle ,
\end{align}
which is what we wanted to prove.

\section{Pseudospectral collocation methods}
\label{a-psc}

We use this appendix to describe very cursorily the PSC methods used for the numerical schemes in this paper and
to introduce some notation in relation thereto. For good detailed expositions see, for example, Refs. \cite{boyd_chebyshev_2001,trefethen_spectral_2001,peyret_chebyshev_2002}.

We
work on Chebyshev-Lobatto (CL) computational grids. On any domain
$[a,b]=\mathscr{D}\subseteq\mathscr{I}$, these comprise the (\emph{non-uniformly}
spaced) set of $N$ points $\{X_{i}\}_{i=0}^{N}\subset\mathscr{D}$
obtained by projecting onto $\mathscr{D}$ those points located at
equal angles on a hypothetical semicircle having $\mathscr{D}$ as
its diameter. That is to say, the CL grid on the ``standard'' spectral
domain $\mathscr{D}^{\text{s}}=[-1,1]$ is given by
\begin{equation}
X_{i}^{\text{s}}=-\cos\left(\frac{\pi i}{N}\right)\,,\quad\forall0\leq i\leq N\,,
\end{equation}
which can straightforwardly be transformed (by shifting and stretching)
to the desired grid on $\mathscr{D}$. For any function $f:\mathscr{D}\rightarrow\mathbb{R}$
we denote via a subscript its value at the $i$-th CL point, $f(X_{i})=f_{i}$,
and in slanted boldface the vector containing all such values,
\begin{equation}
\bm{f}=\left[\begin{array}{c}
f_{0}\\
f_{1}\\
\vdots\\
f_{N}
\end{array}\right]\,.\label{eq:f_cl_grid}
\end{equation}
There exists an $(N+1)\times(N+1)$ matrix $\mathbb{D}$, the so-called
CL differentiation matrix, such that the derivative values of $f$
can be approximated simply by applying it to (\ref{eq:f_cl_grid}),
i.e. $\bm{f}'=\mathbb{D}\bm{f}$. For convenience, we also employ
the notation $\mathbb{M}(r_{\text{i}}:r_{\text{f}},c_{\text{i}}:c_{\text{f}})$
to refer to the part of any matrix $\mathbb{M}$ from the $r_{\text{i}}$-th
to the $r_{\text{f}}$-th row and from the $c_{\text{i}}$-th to the
$c_{\text{f}}$-th column. (A simple ``$:$'' indicates taking all
rows/columns.)

\section{Numerical schemes for distributionally-sourced PDEs}
\label{a-numerical-schemes}

\subsection{First-order hyperbolic PDEs}
\label{a-numerical-first-order-hyperbolic}

We apply a first order in time finite difference scheme to the homogeneous
PDEs; thus, prior to imposing BCs/JCs, the equations become $\frac{1}{\mathrm{\Delta} t}(\bm{u}_{k+1}^{\pm}-\bm{u}_{k}^{\pm})=-\mathbb{D}^{\pm}\bm{u}_{k}^{\pm}$,
where the vectors $\bm{u}_{k}^{\pm}$ contain the values of the solutions
on the CL grids at the $k$-th time step, $\mathbb{D}^{\pm}$ is the
CL differentiation matrix on the respective domains, and $\mathrm{\Delta} t$
is our time step. We can rewrite the discretized PDE as $\bm{u}_{k+1}^{\pm}=\bm{u}_{k}^{\pm}-\mathrm{\Delta} t\mathbb{D}^{\pm}\bm{u}_{k}^{\pm}=\bm{s}_{k}^{\pm}$.
To impose the BC and JC, we modify the equations as follows: 
\begin{equation}
\left[\begin{array}{c}
\bm{u}_{k+1}^{-}\\
\hline \bm{u}_{k+1}^{+}
\end{array}\right]=\left[\begin{array}{c}
u_{N,k}^{+}\\
\bm{s}_{k}^{-}(2:N+1)\\
\hline u_{N,k}^{-}+g_{k}\\
\bm{s}_{k}^{+}(2:N+1)
\end{array}\right].\label{eq:advection_psc}
\end{equation}

Similarly, for our neuroscience application, we discretize the PDE using a first order finite difference scheme:
$\frac{1}{\mathrm{\Delta} t}(\bm{\rho}_{k+1}^{\pm}-\bm{\rho}_{k}^{\pm})=-\mathbb{D}^{\pm}\bm{R}_{k}^{\pm}$
where $R_{i,k}^{\pm}=(1-V_{i}^{\pm})\rho_{i,k}^{\pm}$. Hence, prior
to imposing the BC/JC, we have $\bm{\rho}_{k+1}^{\pm}=\bm{\rho}_{k}^{\pm}-\mathrm{\Delta} t\mathbb{D}^{\pm}\bm{R}_{k}^{\pm}=\bm{s}_{k}^{\pm}$.
To impose the BC/JC, we just modify the equations accordingly:
\begin{equation}
\left[\begin{array}{c}
\bm{\rho}_{k+1}^{-}\\
\hline \bm{\rho}_{k+1}^{+}
\end{array}\right]=\left[\begin{array}{c}
0\\
\bm{s}_{k}^{-}(2:N+1)\\
\hline \rho_{N,k}^{-}+\frac{1-L}{1-V_{*}}\rho_{N,k}^{+}\\
\bm{s}_{k}^{+}(2:N+1)
\end{array}\right].\label{eq:neural_population_psc}
\end{equation}

\subsection{Parabolic PDEs}
\label{a-numerical-parabolic}

In these problems, we have moving boundaries for the CL grids (since
the location of the singular source is time-dependent). The mapping
for transforming the standard (fixed) spectral domain $[-1,1]$ into
an arbitrary (time-dependent) one, say $\mathscr{D}=[a(t),b(t)]$,
is given by
\begin{align}
\mathscr{V}\times\left[0,1\right]\rightarrow\, & \mathscr{V}\times\mathscr{D}\\
\left(T,X\right)\mapsto\, & \left(t\left(T\right),x\left(T,X\right)\right)\,,
\end{align}
where
\begin{align}
t\left(T\right)=\, & T\,,\\
x\left(T,X\right)=\, & \frac{b-a}{2}X+\frac{a+b}{2}\,.
\end{align}
For transforming back, we have
\begin{align}
\mathscr{V}\times\mathscr{D}\rightarrow\, & \mathscr{V}\times\left[0,1\right]\\
\left(t,x\right)\mapsto\, & \left(T\left(t\right),X\left(t,x\right)\right)\,,
\end{align}
where
\begin{align}
T\left(t\right)=\, & t\,,\label{eq:spectral_T(t)}\\
X\left(t,x\right)=\, & \frac{2x-a-b}{b-a}\,.\label{eq:spectral_X(t,x)}
\end{align}
Thus, for any function $f(t,x)$ in these problems, we must take care
to express the time partial using the chain rule as
\begin{align}
\frac{\partial f}{\partial t}=\, & \frac{\partial f}{\partial T}\frac{\partial T}{\partial t}+\frac{\partial f}{\partial X}\frac{\partial X}{\partial t}\\
=\, & \frac{\partial f}{\partial T}-\frac{2}{\left(b-a\right)^{2}}\left[\left(b-x\right)\dot{a}+\left(x-a\right)\dot{b}\right]\frac{\partial f}{\partial X}\,,\label{eq:spectral_partial_t_f}
\end{align}
where in the second line we have used (\ref{eq:spectral_T(t)})-(\ref{eq:spectral_X(t,x)}). 

Now, let us use this to formulate the numerical schemes for our problems---first,
for the heat equation. Let $\mathbb{D}_{k}^{\pm}$ denote the CL differentiation
matrices on each of the two domains at the $k$-th time step. Then,
using (\ref{eq:spectral_partial_t_f}), we have here the following
finite difference formula for the homogeneous PDEs prior to imposing
BCs/JCs: $\frac{1}{\mathrm{\Delta} t}(\bm{u}_{k+1}^{\pm}-\bm{u}_{k}^{\pm})=(\mathbb{D}_{k}^{\pm})^{2}\bm{u}_{k}^{\pm}-\mathbb{C}_{k}^{\pm}\mathbb{D}\bm{u}_{k}^{\pm}$,
where $\mathbb{D}$ is the CL differentiation matrix on $[-1,1]$
and $\mathbb{C}_{k}^{-}={\rm diag}([2/(x_{p}(t_{k}))^{2}][(-x_{i}^{-})\dot{x}_{p}(t_{k})])$,
$\mathbb{C}_{k}^{+}={\rm diag}([2/(1-x_{p}(t_{k}))^{2}][(x_{i}^{+}-1)\dot{x}_{p}(t_{k})])$.
Thus $\bm{u}_{k+1}^{\pm}=\bm{u}_{k}^{\pm}+\mathrm{\Delta} t[(\mathbb{D}_{k}^{\pm})^{2}-\mathbb{C}_{k}^{\pm}\mathbb{D}]\bm{u}_{k}^{\pm}=\bm{s}_{k}^{\pm}$.
We can implement the BCs and JCs, by modifying the first and last
equations on each domain: 
\begin{equation}
\left[\begin{array}{cccc|cccc}
1 & 0 & \cdots & 0 & 0\\
0 & 1 & \cdots & 0 &  & 0\\
\vdots & \vdots & \ddots & \vdots &  &  & \ddots\\
0 & 0 & \cdots & 1 &  &  &  & 0\\
\hline 0 &  &  &  &  &  & \mathbb{D}_{k}^{+}(1,:)\\
 & 0 &  &  & 0 & 1 & \cdots & 0\\
 &  & \ddots &  & \vdots & \vdots & \ddots & \vdots\\
 &  &  & 0 & 0 & 0 & \cdots & 1
\end{array}\right]\left[\begin{array}{c}
\\
\bm{u}_{k+1}^{-}\\
\\
\hline \\
\bm{u}_{k+1}^{+}\\
\\
\end{array}\right]=\left[\begin{array}{c}
0\\
\bm{s}_{k}^{-}(2:N)\\
u_{0,k}^{+}\\
\hline \mathbb{D}_{k}^{-}(N,:)\bm{u}_{k}^{-}-\lambda\\
\bm{s}_{k}^{+}(2:N)\\
0
\end{array}\right].\label{eq:heat_psc}
\end{equation}
Note that we are actually introducing an error by using (for convenience
and ease of adaptability) $\mathbb{D}_{k}^{+}$ instead of $\mathbb{D}_{k+1}^{+}$
on the LHS (in the equation for $u_{0,k+1}^{+}$). However, one can
easily convince oneself that $\mathbb{D}_{k+1}^{+}-\mathbb{D}_{k}^{+}=\mathcal{O}(\mathrm{\Delta} t)$,
which is already the order of the error of the finite difference scheme,
so we are not actually introducing any new error in this way. Furthermore,
because we use up the last equation for $\bm{u}_{k}^{-}$ to impose
the JC on $u$ (i.e. we do not have an equation for $u_{N,k}^{-}$),
we must use the derivative at the previous point (i.e., at $u_{N-1,k}^{-}$)
in order to impose the derivative JC. Hence on the RHS, we use $\mathbb{D}_{k}^{-}(N,:)$
instead of $\mathbb{D}_{k}^{-}(N+1,:)$.

The scheme for the finance model is analogous. We use again the first-order
finite-difference method for the homogeneous equations, $\frac{1}{\mathrm{\Delta} t}(\bm{f}_{k+1}^{\sigma}-\bm{f}_{k}^{\sigma})=(\mathbb{D}_{k}^{\sigma})^{2}\bm{f}_{k}^{\sigma}-\mathbb{C}_{k}^{\sigma}\mathbb{D}\bm{f}_{k}^{\sigma}$
with the matrices $\mathbb{C}_{k}^{\sigma}$ defined similarly to
those in the heat equation problem (again using (\ref{eq:spectral_partial_t_f}));
thus $\bm{f}_{k+1}^{\sigma}=\bm{f}_{k}^{\sigma}+\mathrm{\Delta} t[(\mathbb{D}_{k}^{\sigma})^{2}-\mathbb{C}_{k}^{\sigma}\mathbb{D}]\bm{f}_{k}^{\sigma}=\bm{s}_{k}^{\sigma}$.
To impose the BCs/JCs, we modify the equations appropriately: 
\begin{align}
\left[\begin{array}{ccccc}
 &  & \mathbb{D}_{k}^{-}(1,:)\\
0 & 1 & \cdots & 0 & 0\\
\vdots & \vdots & \ddots & \vdots & \vdots\\
0 & 0 & \cdots & 1 & 0\\
0 & 0 & \cdots & 0 & 1
\end{array}\right]\left[\begin{array}{c}
\\
\\
\bm{f}_{k+1}^{-}\\
\\
\\
\end{array}\right] & =\left[\begin{array}{c}
0\\
s_{1,k}^{-}\\
\vdots\\
s_{N,k}^{-}\\
f_{0,k}^{0}
\end{array}\right],\label{eq:finance_psc_minus}\\
\left[\begin{array}{ccccc}
 &  & \mathbb{D}_{k}^{0}(1,:)\\
0 & 1 & \cdots & 0 & 0\\
\vdots & \vdots & \ddots & \vdots & \vdots\\
0 & 0 & \cdots & 1 & 0\\
0 & 0 & \cdots & 0 & 1
\end{array}\right]\left[\begin{array}{c}
\\
\\
\bm{f}_{k+1}^{0}\\
\\
\\
\end{array}\right] & =\left[\begin{array}{c}
\mathbb{D}_{k}^{-}(N,:)\bm{f}_{k}^{-}-\lambda_{k}\\
s_{1,k}^{0}\\
\vdots\\
s_{N,k}^{0}\\
f_{0,k}^{+}
\end{array}\right],\label{eq:finance_psc_zero}\\
\left[\begin{array}{ccccc}
 &  & \mathbb{D}_{k}^{+}(1,:)\\
0 & 1 & \cdots & 0 & 0\\
\vdots & \vdots & \ddots & \vdots & \vdots\\
0 & 0 & \cdots & 1 & 0\\
 &  & \mathbb{D}_{k}^{+}(N+1,:)
\end{array}\right]\left[\begin{array}{c}
\\
\\
\bm{f}_{k+1}^{+}\\
\\
\\
\end{array}\right] & =\left[\begin{array}{c}
\mathbb{D}_{k}^{0}(N,:)\bm{f}_{k}^{0}+\lambda_{k}\\
s_{1,k}^{+}\\
\vdots\\
s_{N,k}^{+}\\
0
\end{array}\right].\label{eq:finance_psc_plus}
\end{align}

\subsection{Second-order hyperbolic PDEs}
\label{a-numerical-second-order-hyperbolic}

We again apply a first order in time finite difference scheme to the
homogeneous PDEs; prior to imposing BCs/JCs, the equations become
\begin{equation}
\frac{1}{\mathrm{\Delta} t}\left(\left[\begin{array}{c}
\bm{u}_{k+1}^{\pm}\\
\bm{v}_{k+1}^{\pm}\\
\bm{w}_{k+1}^{\pm}
\end{array}\right]-\left[\begin{array}{c}
\bm{u}_{k}^{\pm}\\
\bm{v}_{k}^{\pm}\\
\bm{w}_{k}^{\pm}
\end{array}\right]\right)=\mathbb{C}^{\pm}\left[\begin{array}{c}
\bm{u}_{k}^{\pm}\\
\bm{v}_{k}^{\pm}\\
\bm{w}_{k}^{\pm}
\end{array}\right]\,,
\end{equation}
where
\begin{equation}
\mathbb{C}^{\pm}=\left[\begin{array}{ccc}
0 & 0 & 0\\
0 & 0 & \mathbb{I}\\
0 & \mathbb{I} & 0
\end{array}\right]\left[\begin{array}{ccc}
\mathbb{D}^{\pm} & 0 & 0\\
0 & \mathbb{D}^{\pm} & 0\\
0 & 0 & \mathbb{D}^{\pm}
\end{array}\right]+\left[\begin{array}{ccc}
0 & 0 & \mathbb{I}\\
0 & 0 & 0\\
0 & 0 & 0
\end{array}\right]=\left[\begin{array}{ccc}
0 & 0 & \mathbb{I}\\
0 & 0 & \mathbb{D}^{\pm}\\
0 & \mathbb{D}^{\pm} & 0
\end{array}\right]\,.
\end{equation}
We can rewrite the discretized PDE as 
\begin{equation}
\left[\begin{array}{c}
\bm{u}_{k+1}^{\pm}\\
\bm{v}_{k+1}^{\pm}\\
\bm{w}_{k+1}^{\pm}
\end{array}\right]=\left(\mathrm{\Delta} t\mathbb{C}^{\pm}+\mathbb{I}\right)\left[\begin{array}{c}
\bm{u}_{k}^{\pm}\\
\bm{v}_{k}^{\pm}\\
\bm{w}_{k}^{\pm}
\end{array}\right]=\left[\begin{array}{c}
\bm{s}_{k}^{\pm}\\
\bm{y}_{k}^{\pm}\\
\bm{z}_{k}^{\pm}
\end{array}\right]\,.
\end{equation}
To impose the BCs and JCs, we modify the equations as follows: 
\begin{align}
\left[\begin{array}{cccc}
1 & \cdots & 0 & 0\\
\vdots & \ddots & \vdots & 0\\
0 & \cdots & 1 & 0\\
 & \mathbb{D}^{-}(N+1,:)
\end{array}\right]\left[\begin{array}{c}
\\
\bm{u}_{k+1}^{-}\\
\\
\end{array}\right]=\, & \left[\begin{array}{c}
0\\
\bm{s}_{k}^{-}(2:N)\\
\mathbb{D}^{+}(1,:)\bm{u}_{k}^{+}
\end{array}\right]\,,\\
\left[\begin{array}{c}
\\
\bm{u}_{k+1}^{+}\\
\\
\end{array}\right]=\, & \left[\begin{array}{c}
u_{N-1,k+1}^{-}-g_{k}\\
\bm{s}_{k}^{+}(2:N)\\
0
\end{array}\right]\,,\\
\left[\begin{array}{c}
\bm{v}_{k+1}^{-}\\
\hline \bm{v}_{k+1}^{+}
\end{array}\right]=\, & \left[\begin{array}{c}
\mathbb{D}^{-}\bm{u}_{k+1}^{-}\\
\hline \mathbb{D}^{+}\bm{u}_{k+1}^{+}
\end{array}\right]\,,\\
\left[\begin{array}{c}
\bm{w}_{k+1}^{-}\\
\hline \bm{w}_{k+1}^{+}
\end{array}\right]=\, & \left[\begin{array}{c}
0\\
\bm{z}_{k}^{-}(2:N+1)\\
\hline w_{N,k+1}^{-}-\dot{g}_{k+1}\\
\bm{z}_{k}^{+}(2:N)\\
0
\end{array}\right]\,,
\end{align}

\subsection{Elliptic PDEs}
\label{a-numerical-elliptic}

In this case we have no time evolution, and we simply need to solve
$((\mathbb{D}^{\pm})^{2}+{\rm diag}(1/X_{i}^{\pm})\mathbb{D}^{\pm})\bm{u}^{\pm}=\mathbb{M}^{\pm}\bm{u}^{\pm}={\bf 0}$,
modified appropriately to account for the BCs and JCs. In particular,
we first solve for $\bm{u}^{+}$ using the BCs, and then for $\bm{u}^{-}$
using the solution for $\bm{u}^{+}$ to implement the JCs:
\begin{align}
\left[\begin{array}{c}
\mathbb{M}^{+}(1:N-1,:)\\
0\enskip0\cdots0\enskip1\\
\mathbb{D}^{+}(N+1,:)
\end{array}\right]\bm{u}^{+}=\, & \left[\begin{array}{c}
{\bf 0}(1:N-1)\\
1-\tfrac{1}{2}\log(2L)\\
-\tfrac{1}{2L}
\end{array}\right]\,,\\
\left[\begin{array}{c}
\mathbb{M}^{-}(1:N-1,:)\\
0\enskip0\cdots0\enskip1\\
\mathbb{D}^{-}(N+1,:)
\end{array}\right]\bm{u}^{-}=\, & \left[\begin{array}{c}
{\bf 0}(1:N-1)\\
u_{0}^{+}\\
\mathbb{D}^{+}(1,:)\bm{u}^{+}+1
\end{array}\right]\,.
\end{align}

\section{Exact solution for the elastic wave equation}
\label{a-solution-elastic-wave}

A useful method for obtaining exact solutions to the problem (\ref{eq:elastic_wave_pde})
on $\mathscr{I}=\mathbb{R}$ is outlined in Ref. \cite{petersson_stable_2010}; we follow
the same procedure here, except using a sinusoidal source time function
(rather than a polynomial, as is done in Ref. \cite{petersson_stable_2010}).

We begin by Fourier transforming the PDE in the spatial domain, using
\begin{equation}
\hat{u}\left(\xi,t\right)=\int_{\mathbb{R}}{\rm d}x\,u\left(x,t\right){\rm e}^{-{\rm i}\xi x}\,,\quad u\left(x,t\right)=\frac{1}{2\pi}\int_{\mathbb{R}}{\rm d}\xi\,\hat{u}\left(\xi,t\right){\rm e}^{{\rm i}\xi x}\,.\label{eq:fourier_transform_definition}
\end{equation}
Thus, multiplying the PDE in (\ref{eq:elastic_wave_pde}) by ${\rm e}^{-{\rm i}\xi x}$,
integrating over $\mathbb{R}$ and applying integration by parts with
the assumption of vanishing boundary terms, we get the following equation
for the Fourier transform of $u$:
\begin{equation}
\ddot{\hat{u}}+\xi^{2}\hat{u}={\rm i}\xi g\left(t\right){\rm e}^{-{\rm i}\xi x_{*}}\,.\label{eq:elastic_wave_equation_ft}
\end{equation}
One can easily check that the exact solution of (\ref{eq:elastic_wave_equation_ft}),
with initial conditions $\hat{u}(\xi,0)=0=\dot{\hat{u}}(\xi,0)$,
is simply
\begin{equation}
\hat{u}\left(\xi,t\right)=\frac{1}{2}{\rm e}^{-{\rm i}\xi x_{*}}\sum_{\sigma=\pm}\sigma{\rm e}^{\sigma{\rm i}\xi t}\int_{0}^{t}{\rm d}\tau\,g\left(\tau\right){\rm e}^{-\sigma{\rm i}\xi\tau}\,.\label{eq:elastic_wave_solution_exact_ft}
\end{equation}
Inserting $g(\tau)=\kappa\sin(\omega\tau)$ into (\ref{eq:elastic_wave_solution_exact_ft})
and carrying out the integrals, we get
\begin{equation}
\hat{u}=\frac{{\rm i}\kappa{\rm e}^{-{\rm i}\xi x_{*}}}{\xi^{2}-\omega^{2}}\left[\xi\sin\left(\omega t\right)-\omega\sin\left(\xi t\right)\right]\,.\label{eq:elastic_wave_solution_exact_ft_2}
\end{equation}
Finally, plugging (\ref{eq:elastic_wave_solution_exact_ft_2}) back
into (\ref{eq:fourier_transform_definition}) and using
\begin{align}
\int_{\mathbb{R}}{\rm d}\xi\,\frac{\xi{\rm e}^{{\rm i}\xi\left(x-x_{*}\right)}}{\xi^{2}-\omega^{2}}=\, & {\rm i}\pi{\rm sgn}\left(x-x_{*}\right)\cos\left(\omega\left(x-x_{*}\right)\right)\,,\\
\int_{\mathbb{R}}{\rm d}\xi\,\frac{\sin\left(\xi t\right){\rm e}^{{\rm i}\xi\left(x-x_{*}\right)}}{\xi^{2}-\omega^{2}}=\, & \frac{{\rm i}\pi}{2\omega}\sum_{\sigma=\pm}\sigma{\rm sgn}\left(x-x_{*}+\sigma t\right)\sin\left(\omega\left(x-x_{*}+\sigma t\right)\right)\,,
\end{align}
we get the solution (\ref{eq:elastic_wave_solution_exact}).



\bibliographystyle{spphys}
\bibliography{main.bbl}

\end{document}